\documentclass{aa}  

\usepackage{graphicx}
\usepackage{booktabs}
\usepackage{natbib}
\bibpunct{(}{)}{;}{a}{}{,} % to follow the A&A style
\usepackage{txfonts,textcomp}
\usepackage{lscape}
\usepackage[allcolors=blue]{hyperref}
\usepackage{cleveref}
\begin{document}

   \title{Central radio galaxies in galaxy clusters: Joint surveys by eROSITA and ASKAP}

%   \subtitle{I. Overviewing the $\kappa$-mechanism}

   \author{K. Böckmann\inst{1}, M. Brüggen\inst{1}, B. Koribalski\inst{2, 3}, A. Veronica\inst{4}, T.H. Reiprich\inst{4}, E. Bulbul \inst{5}, Y. E. Bahar\inst{5}, F. Balzer\inst{5}, J. Comparat\inst{5}, C. Garrel\inst{5}, V. Ghirardini\inst{5}, G. Gürkan\inst{6,7,8}, M. Kluge\inst{5}, D. Leahy\inst{9}, A. Merloni\inst{5}, A. Liu\inst{5}, M. E. Ramos-Ceja\inst{5}, M. Salvato\inst{5}, J. Sanders\inst{5},  S. Shabala\inst{10}, X. Zhang\inst{5}}

\authorrunning{K. Böckmann et al.}

   \institute{Hamburger Sternwarte, Universität Hamburg, Gojenbergsweg 112, 21029 Hamburg
        \and
        Australia Telescope National Facility, CSIRO Astronomy and Space Science, P.O. Box 76, Epping, NSW 1710, Australia
        \and 
        School of Science, Western Sydney University, Locked Bag 1797, Penrith, NSW 2751, Australia
        \and
        Argelander Institute for Astronomy (AIfA), University of Bonn, Auf dem Hügel 71, 53121 Bonn, German
        \and 
        Max-Planck-Institut für extraterrestrische Physik, Gießenbachstraße 1, D-85748 Garching, Germany  
        \and 
        Center for Astrophysics Research, Department of Physics, Astronomy and Mathematics, University of Hertfordshire, College Lane, Hatfield AL10 9AB, UK
        \and
        Thüringer Landessternwarte, Sternwarte 5, D-07778 Tautenburg, Germany
        \and 
        CSIRO Space and Astronomy, ATNF, PO Box 1130, Bentley WA 6102, Australia
        \and
        Department of Physics and Astronomy, University of Calgary, Calgary, Alberta, Canada T2N 1N4 
        \and
        School of Natural Sciences, University of Tasmania, Private Bag 37, Hobart, Tasmania 7001, Australia
        }

   \date{04/13/2023}

% \abstract{}{}{}{}{} 
% 5 {} token are mandatory
 
  \abstract
   {The extended ROentgen Survey with an Imaging Telescope Array (eROSITA) telescope on board the Spectrum-Roentgen-Gamma (SRG) mission has completed the first eROSITA All-Sky Survey (eRASS:1). It detected $\sim 10^4$ galaxy clusters in the western Galactic hemisphere. In the radio band, the Australian Square Kilometre Array Pathfinder (ASKAP) telescope completed its pilot 1 phase of the project Evolutionary Map of the Universe (EMU) with 220.000 sources in a 270 deg$^2$ field overlapping with eRASS:1. These two surveys are used to study radio-mode active galactic nuclei in clusters.}
   {In order to understand the efficiency of radio-mode feedback at the centers of galaxy clusters, we relate the radio properties of the brightest cluster galaxies to the X-ray properties of the host clusters.}
   {We identified the central radio sources in eRASS:1 clusters or calculated corresponding upper limits on the radio luminosity. Then, we derived relations between the X-ray properties of the clusters and the radio properties of the corresponding central radio source.} %We also apply a mid-infrared color criterion using WISE colors to identify AGN.}
   {In total, we investigated a sample of 75 clusters. We find a statistically significant correlation between the X-ray luminosity of the cluster and the 944 MHz radio luminosity of the corresponding central radio galaxy. There is also a positive trend between the radio power and the largest linear size of the radio source. The density and the largest linear size are not correlated. We find that the kinetic luminosity of the radio jets  in high-luminosity clusters
with $L_{\mathrm{X}} > 10^{43}$ erg s$^{-1}$ is no longer correlated with the X-ray luminosity, and we discuss various reasons. We find an anticorrelation between the central cooling time $t_{\mathrm{cool}}$ and the radio luminosity $L_{\mathrm{R}}$ , indicating a need for more powerful active galactic nuclei in clusters with short central cooling times.} %The relation between kinetic luminosity and X-ray luminosity suggests that heating from the central AGN may counterbalance ICM radiative losses in most clusters. }
   {}


   \maketitle
%
%------------------------------------------------------------------- Introduction
\section{Introduction}
The hot gas in galaxy clusters, called the intracluster medium (ICM), is strongly affected by feedback of active galactic nuclei (AGN). AGN reside in the dominant cD galaxies of clusters, which host the most massive black holes of the Universe. Two different modes of AGN feedback are found in observations: the radiative or quasar mode, and the radio or jet mode \citep{Cattaneo_2009, Heckman_2014}. The radiative mode causes uniform heating of the environment around it, whereas in radio mode, AGN jets expel radio-heated gas from the accreting black hole matter into the ICM and push the existing X-ray heated cluster gas away \citep{Shabala_2020}. Therefore, on the one hand, AGN activity prevents the cooling of gas and subsequent star formation, and on the other hand, AGN contribute to star formation by projecting jets toward the ICM and compressing the gas. An increment of radiative losses of the ICM in turn leads to an increment in heating of the gas by the AGN. The more gas cools, the higher the energy output that is able to quench the radiative losses. This is known as the \emph{AGN feedback loop\/} \citep{McNamara_2016, Gaspari_2020}. AGN feedback has been observed in a wide range of systems, from isolated elliptical galaxies to massive clusters. The most powerful AGN operating in radio mode can be found in the brightest cluster galaxies (BCGs), which usually are massive elliptical galaxies residing at the bottom of the clusters potential (e.g., reviews by \citet{Fabian_2012}, \citet{Gitti_2012}, \citep{McNamara_2012}). \\

\citet{Ineson_2013, Ineson_2015} performed a study of the interplay between AGN and their environment in a sample of radio-loud AGN galaxy clusters and galaxy groups. As a result, they found a correlation between the X-ray emission from the ICM and the power of the corresponding central radio source at 151 MHz. They argued that this correlation could arise from AGN in a phase of radiatively inefficient accretion, which are also called low-excitation radio galaxies (LERGs), while high-excitation radio galaxies (HERGs) stand out of the distribution and show higher radio powers. However, it is important to note that the origin of this relation is not obvious as X-ray emission in clusters and groups is mostly due to line emission and Bremsstrahlung that allows the ICM to cool. Therefore, the timescale of these radiative losses is strongly dependent on the distance of the diffuse gas from the cluster core, which varies from less than 1 Gyr in the center of the strongest cool-cores to a few billion years in the outskirts. \citet{Nipoti}, on the other hand, suggested that the AGN power output reoccurs and acts in cycles of $\sim 10^8$ yr. As a consequence, the timescales of these two processes are usually significantly different. On the other hand, \citet{Hardcastle_2019} suggested that the majority of sources are compact and therefore short-lived, and so the cycle is many short-lived ($\sim$1 Myr) intermittent bursts (see also \citep{Shabala_2008}). \citet{Pasini_2020, Pasini_2021} and \citet{Pasini_2022} investigated the radio power of the central radio galaxy in galaxy clusters and in galaxy groups at different wavelengths. The authors also found positive correlations between the radio power of the central AGN at frequencies from 144 MHz to 1.4 GHz and the main properties of the diffuse X-ray intragroup and intracluster medium, again suggesting a link between AGN heating and cooling processes in the gaseous halo \citep{Pope_2012}.\\

Deep X-ray observations with \emph{XMM-Newton\/} and \emph{Chandra\/} have revealed that most of the systems with radio-mode AGN have disturbed X-ray morphologies caused by AGN-ejected jets. These surface brightness features, including cavities in the X-ray images and sharp density discontinuities that are interpreted as shocks, indicate a strong correlation between the ICM and the central AGN. The thermodynamical properties of the intracluster gas is also affected by AGN feedback in teRMS of the gas entropy distribution and transport of high-metallicity gas from the center of the cluster to its outskirts. The X-ray cavities or bubbles that have been discovered in X-ray images of clusters are often filled with radio emission. This leads to the assumption that radio plasma produced by AGN outflows displaces the X-ray emitting gas of the ICM. One of the main results of these observations was the revelation of a scaling relation between the cavity power and the radio luminosity \citep{Birzan_2004, Rafferty_2006, Birzan_2020, Timmermann_2020}. \\
The extended ROentgen Survey with an Imaging Telescope Array (eROSITA) on board the Spectrum-Roentgen-Gamma (SRG) mission was launched in July 2019 \citep{Predehl_2021}. 
eROSITA will perform all-sky surveys (eRASS) with a significantly improved sensitivity compared to the ROSAT all-sky survey. In contrast to X-ray telescopes such as XMM-Newton or Chandra, which are used for long-exposure pointed observations of single targets, eROSITA allows unique survey science capabilities by scanning large areas of the X-ray sky fast and efficiently. The eRASS survey is detecting a large number of previously undetected galaxy clusters and will substantially extend existing galaxy cluster catalogs \citep{Merloni_2012, Liu_2022}. \\

In this work, we make use of the cluster catalog resulting from the first all-sky scan (eRASS:1), which was completed in 2020 (Bulbul et al. (2023, in prep.)). We study the central radio galaxies in the cluster centers with 944 MHz radio data from the survey called Evolutionary Map of the Universe (EMU) that was performed by the Australian Square Kilometre Array Pathfinder (ASKAP) \citep{Norris_2021}.\\

The paper is structured as follows: In \cref{The_Data_sec} we explain how our sample was created, and we display its main properties. In \cref{Analysis_sec} we examine and discuss the correlation between the radio and the X-ray emission from our sample in comparison to other correlations. Finally, we conclude in \cref{Conclusion_sec}. Throughout this paper, we assume the standard $\Lambda$CDM cosmology with H$_0$ = 70 km s$^{-1}$ Mpc$^{-1}$, $\Omega_{\Lambda} = 0.7,$ and $\Omega_{\mathrm{M}} = 0.3$.

%-------------------------------------------------------------------- Data
\section{Data}
\label{The_Data_sec}
\subsection{eRASS:1 cluster catalog}
eROSITA is an X-ray space telescope operating in the 0.2 - 10 keV energy range on board the SRG mission \citep{Sunyaev_2021} (for more information on this mission, see, e.g., the instrument paper by \citet{Predehl_2021}). It has an effective area of 1365cm$^2$ and a spectral resolution of ~80 eV FWHM (Full Width at Half Maximum) at 1 keV and an angular survey resolution of 26 arcseconds. The first task of eROSITA is to scan the whole X-ray sky with a final depth of about 1.3ks. The sensitivity will therefore be improved by at least a factor of 20 compared to the only previous X-ray all-sky survey performed by ROSAT 30 years ago. The main task of eROSITA is the study of evolution and nature of dark energy. eROSITA is expected to detect about 10$^{5}$ galaxy clusters and more than one million AGN \citep{Merloni_2012,Merloni_2020}.\\
The first all-sky survey eRASS:1 imaged the whole X-ray sky over the course of 182 days from December 2019 to June 2020, with an average effective exposure of 150-200s. About $10^4$ clusters are detected as extended sources in eRASS:1 using the source detection algorithm in the eROSITA Standard Analysis Software System (eSASS; \citep{Brunner_2022}). Redshifts are determined using data from the Legacy Survey. Details about the eRASS:1 galaxy cluster catalog can be found in Bulbul et al. (2023). The X-ray luminosity that is used throughout this paper was calculated in the 0.2-2.3 keV band.

\subsection{EMU pilot field}

The Australian Square Kilometer Array Pathfinder (ASKAP) is a radio telescope in the Murchison region of Western Australia \citep{Johnston_2008, Hotan_2021, Koribalski_2022_2}. ASKAP is a radio interferometer consisting of 36 12-meter dish antennas, spread out in two dimensions with baselines up to 6~km. Each antenna is equipped with a wide-field phased array feed (PAF) that is used to form 36 beams, that is, each pointing reaches a field of view of $\sim$30 deg$^2$. \\
The survey project called evolutionary map of the Universe (EMU) uses the ASKAP telescope. In July to August 2019, EMU observed a pilot field for 100 hours to test the planned observing mode for the full EMU survey. The EMU pilot survey mapped 270 deg$^2$ of sky with a RA from 305° to 335° and DEC from -62° to -48,° centered at 944 MHz down to an RMS of about 25-30 $\mu$Jy/beam at an angular resolution of 10 - 18 arcsec (see \citet{Norris_2011, Norris_2021} for further details on the survey). \\
As images with large fields of view such as the EMU pilot field yield a large number of detected astronomical sources, an automated source detection technique that measures the properties of the sources is essential. The most common approach is to identify local peaks of emission above some threshold, and fitting two-dimensional Gaussians. As radio surveys have become deeper and wider in recent years, the number of sources in catalogs has grown enormously, such that manual source-finding and identification is no longer feasible. An ASKAP/EMU source-finding challenge by \citet{Hopkins_2015} that was carried out before the start of EMU tested several approaches for an automatic source detection. Source-finding and cataloging is the last step in the ASKAP data-processing pipeline (ASKAPsoft; e.g., \citet{Guzmann_2019, Wieringa_2020}) for all surveys such as continuum, HI emission and absorption, and polarization, carried out using Selavy \citep{Whiting_2017}. More powerful source finders will likely be applied by each team to their specific projects. \\
The image data were first processed by the ASKAPsoft pipeline, and the subsequent source extraction of the final calibrated image used the software tool Selavy. This tool identifies radio islands with emissions higher than five times the local RMS, and it fits Gaussians to peaks of emission within the islands. The peak as well as the integrated radio flux of each island is computed and stored in a FITS catalog containing a total number of $\sim$220.000 radio islands, $\sim$180.000 of which are single-component sources. In comparison with previous surveys, EMU explores a novel region of parameter space because of the ASKAP wide field of view combined with high angular resolution as well good sensitivity.
%Other source finding methods are also used, e.g., for Wallaby HI cubes \citep{Koribalski_2020} where the algorithm SoFiA was applied which is optimized to work on 3D cubes \citep{Serra_2015, Westmeier_2021}. Also machine learning techniques have been applied to the survey data to find and to classify sources \citep{Gupta_2022}. 
For a summary of the EMU pilot survey specifications, see \Cref{EMU_Spec}.
%--------------------------------------------------- EMU Spec Table
\begin{table}[!h]
%\begin{center}
\caption[]{EMU pilot survey specifications}
\label{EMU_Spec}
\begin{tabular}{ll}
\hline
Area of survey & 270 deg$^2$  \\
Synthesised beamwidth & 13 arcsec $\times 11$ arcsec FWHM \\
Frequency range & 800 -- 1088 MHz \\
RMS sensitivity & 25--35 $\mu$Jy / beam \\
Total integration time & 10 $\times$ 10 hours \\
Number of sources & $\sim$ 200,000 \\
\hline
\end{tabular}
\label{specs}
%\medskip\\
%\end{center}
\end{table}
%----------------------------------------------------------------- 

\subsection{Construction of the sample}

The EMU pilot field is fully covered by eRASS:1. Therefore, we created a cluster sample with all eRASS:1 detected clusters within the EMU field, resulting in a total number of 75 confirmed eRASS:1 clusters. Each cluster was visually inspected in the EMU image as well as in WISE and in legacy optical data to identify the BCG of each cluster and the corresponding radio source to the BCG. \\
We identified the BCG and the corresponding central radio island for each cluster. For 64 of the 75 clusters, we found a central radio source within a distance of $\sim \theta$ to the BCG, where  $\theta$ = 18 arcsec, which is the synthesized beam of the radio observation. For the remaining 11 clusters, we set an upper limit of $3 \sigma$, where $\sigma$ is the RMS noise of the EMU image of RMS = 35 $\mu$Jy. \citet{Magliocchetti_2007} examined radio emission of 550 X-ray selected clusters and reported that only 27\% host a central radio source. However, the difference in these results can be attributed to the depth of the respective datasets because the depth of the survey they used reaches only 3 mJy, whereas the EMU has a depth of 35 $\mu$Jy.\\

Visual inspection of each source is the most reliable way to minimize the number of false identifications because at all separations, some radio identifications selected by the position offset alone will be random coincidences \citep{Condon_2002, Sadler_2002, Mauch_2007}. Nonetheless, we still expect a fraction of false associations, which we describe via the $P$-statistics. This quantifies the probability that a radio source has a chance coincidence within a distance $\theta$ from a certain point, here our BCG candidate. It is given by
\begin{align}
P(\theta) = 1 - e^{n \pi^2} ,
\end{align}
with $n$ denoting the number density of radio sources \citep{Scott_2008}. When we assume a uniform distribution of radio sources of $n = 815$ $\mathrm{ deg}^{-2}$, which is the average source density from an ASKAP observation, we obtain $P(18 \mathrm{arcsec}) = 6.6\%$  contamination. For the 64 clusters with a radio match, we therefore expect about three false associations.

\subsection{Properties of the sample}

\begin{figure}[h]
   \centering
   \includegraphics[width=9.5cm]{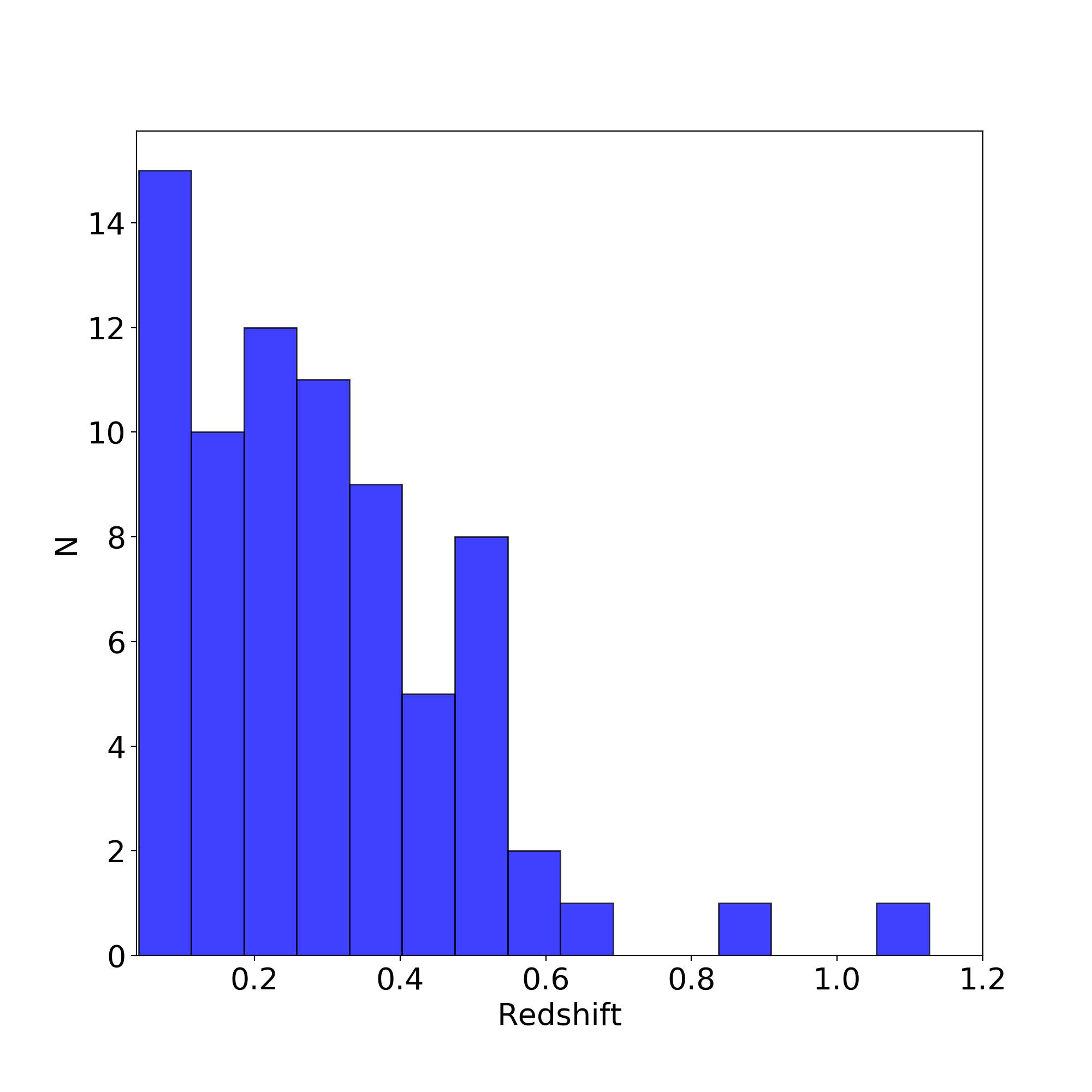}
   \includegraphics[width=9.5cm]{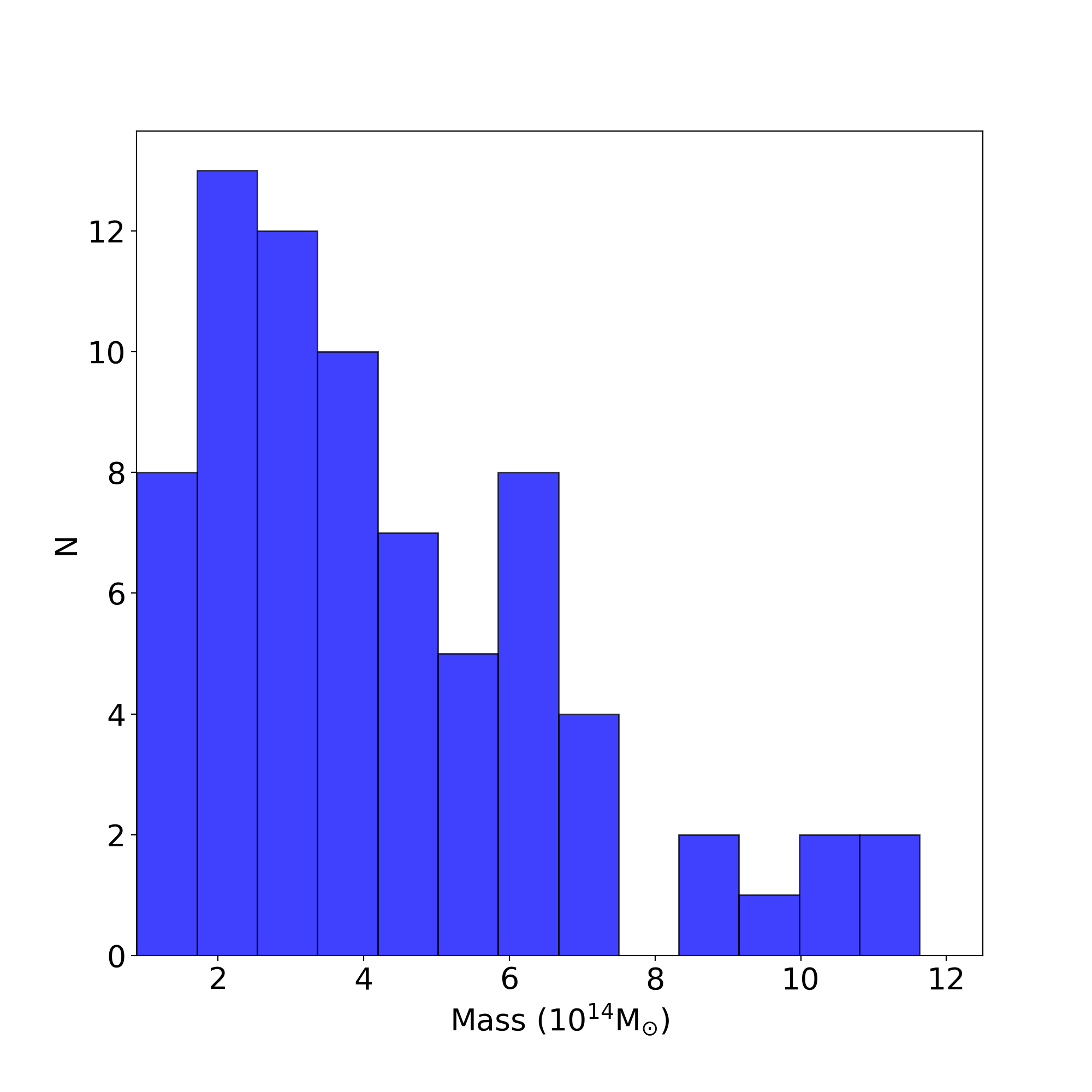}
   \caption{Redshift and mass distribution of all clusters. \emph{Upper panel:\/} Histogram showing the redshift distribution of the sample. \emph{Lower panel:\/} Mass distribution of M$_{500}$ of the cluster sample. The mass was estimated via the L$_\mathrm{X}$ - M$_{500}$ correlation from \citet{Chiu_2022}.}
         \label{Number_Density}
   \end{figure}

\begin{figure}[h]
   \centering
   \includegraphics[width=9cm]{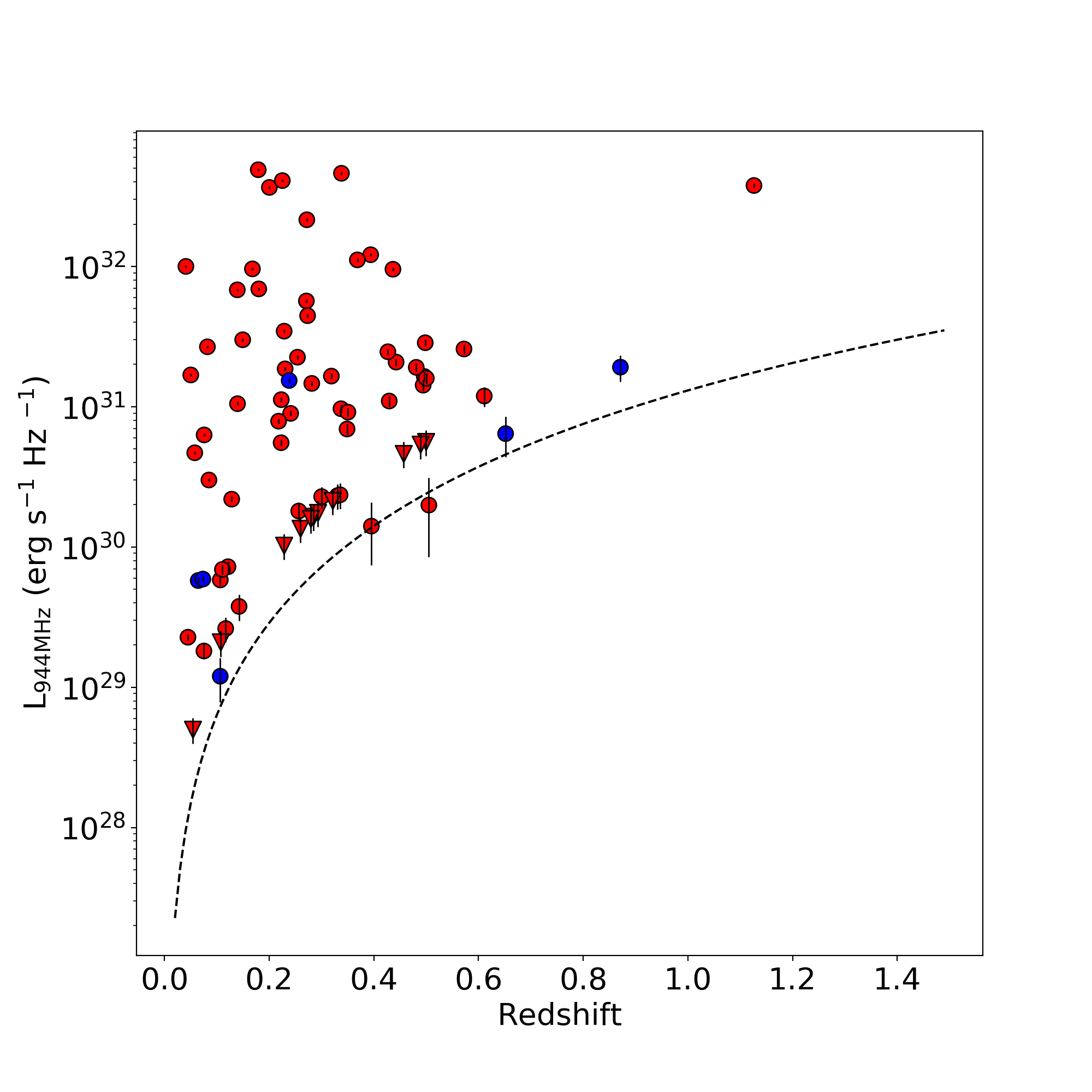}
   \caption{944 MHz radio luminosity vs. redshift. The dashed line represents the theoretical flux limit for point sources. The circles represent the clusters with a detected central radio source, and the triangles represent the upper limits. Resolved sources are plotted in red, and the blue points represent point sources.}
         \label{L_R_z}
   \end{figure}

The upper panel of \Cref{Number_Density} shows a histogram of the redshift distribution of our cluster sample. The majority of the clusters lie within a redshift range of $0.1 < z < 0.7.$  Two outliers lie above $z >0.8$. We used the best available redshift provided from the eRASS:1 cluster catalog, which can be spectroscopic or photometric redshifts. The lower panel of \Cref{Number_Density} shows the mass distribution. The mass was estimated via the L$_\mathrm{X}$ - M$_{500}$ correlation by \citet{Chiu_2022}. The masses of most of the clusters lie within 1 - 12 $\cdot$ 10$^{14}$ M$_{\odot}$. \\
%----------------------------------------------------------------- Number Density z and M
%----------------------------------------------------------------- 
The luminosity of all radio sources was calculated including the following $k$-correction:

\begin{align}
    L_{\mathrm{Radio}} = 4 \pi D_{\mathrm{L}}^2 S_{\mathrm{Radio}}(1+z)^{\alpha - 1}.
\end{align}
$D_{\mathrm{L}}$ is the luminosity distance at redshift $z$ , and $\alpha$ is the spectral index, assumed to be  0.6. \Cref{L_R_z} shows the radio luminosity distribution at 944 MHz versus the redshift of the sample. We also plot the theoretical flux cut in \Cref{L_R_z}. \\

\begin{figure}[h]
   \centering
   \includegraphics[width=9cm]{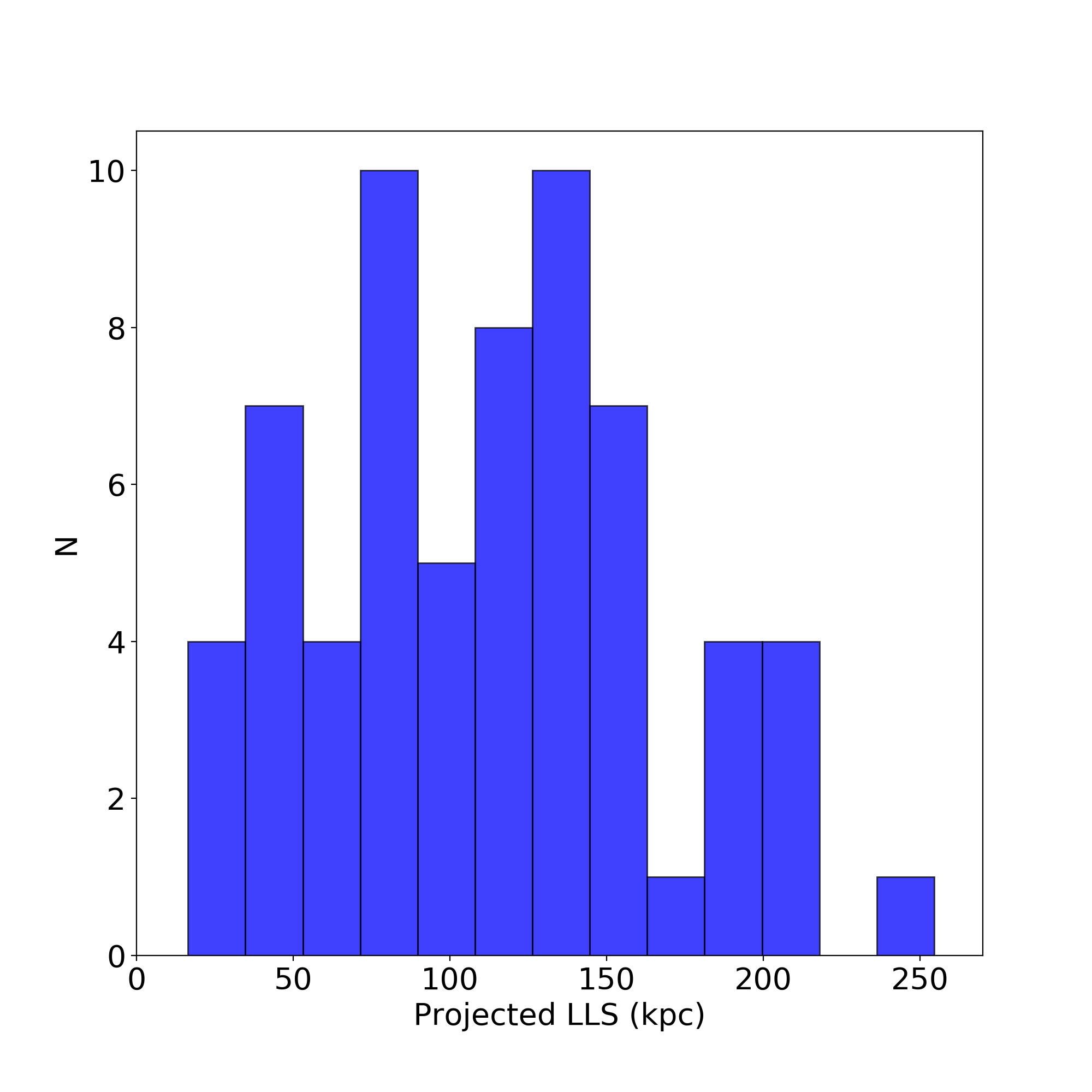}
   \includegraphics[width=9cm]{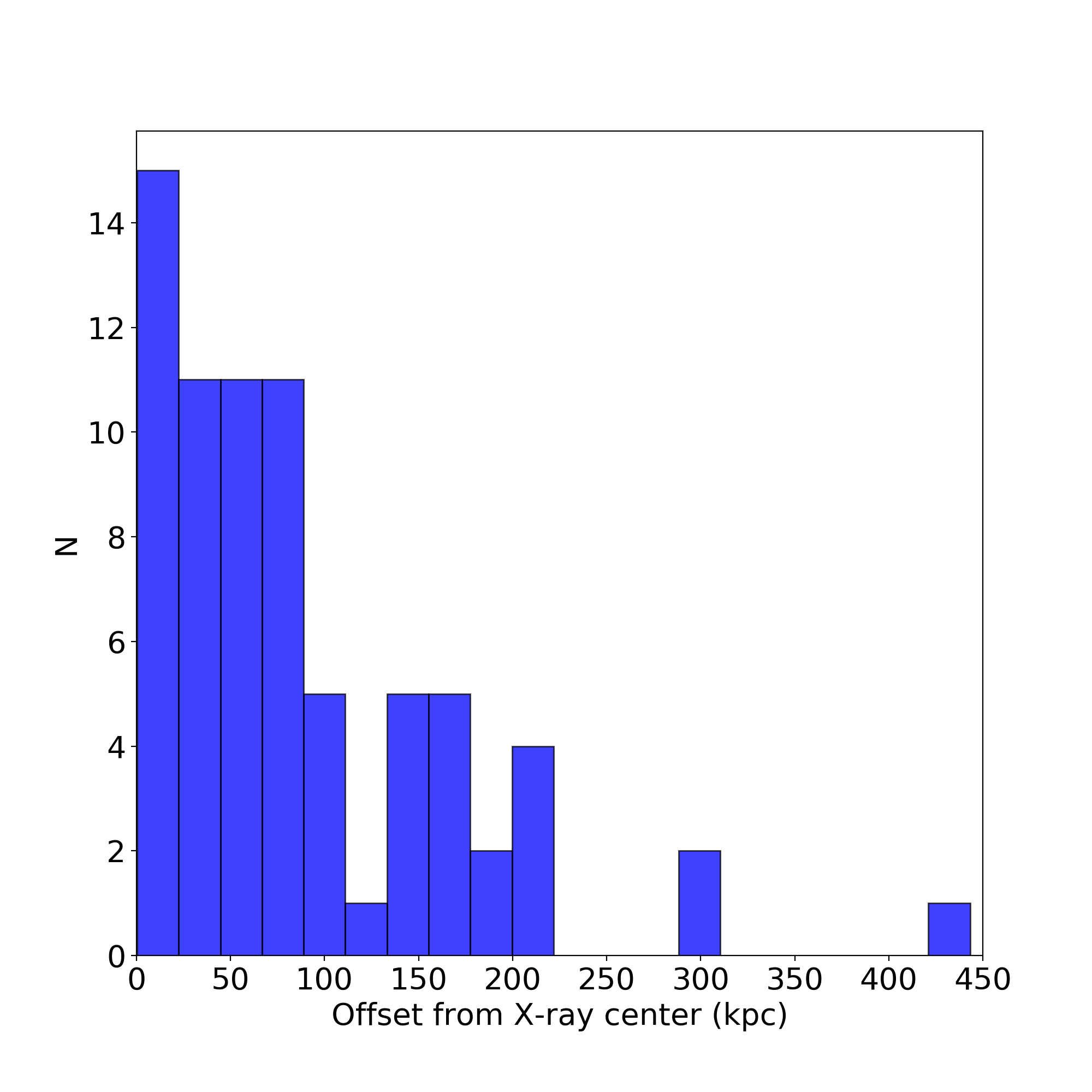}
   \caption{Characteristics of the central radio sources. \emph{Upper panel:\/} Histogram showing the LLS in kiloparsec of each extended radio source. \emph{Lower panel: \/}Histogram showing the BCG offsets from the X-ray emission peak of the cluster.}
         \label{N_Offset}
   \end{figure}

Next, we calculated the largest linear size (LLS) for each radio source. The LLS is defined as the linear size of the major axis of a source, and it is displayed in \Cref{N_Offset}. The LLS is calculated within a 3$\sigma$ isophote. When the radio source was not resolved, which was the case for six clusters, we treated the source as an upper limit with an LLS corresponding to the beam size. The LLS varied from $\sim$ 50 to $\sim$ 250 kpc.  We also show the offset from each BCG to the X-ray center of the corresponding cluster. The majority of BCGs are found within 200 kpc around the X-ray peak of the cluster. \\

\begin{figure*}
   \centering
   \includegraphics[width=9cm]{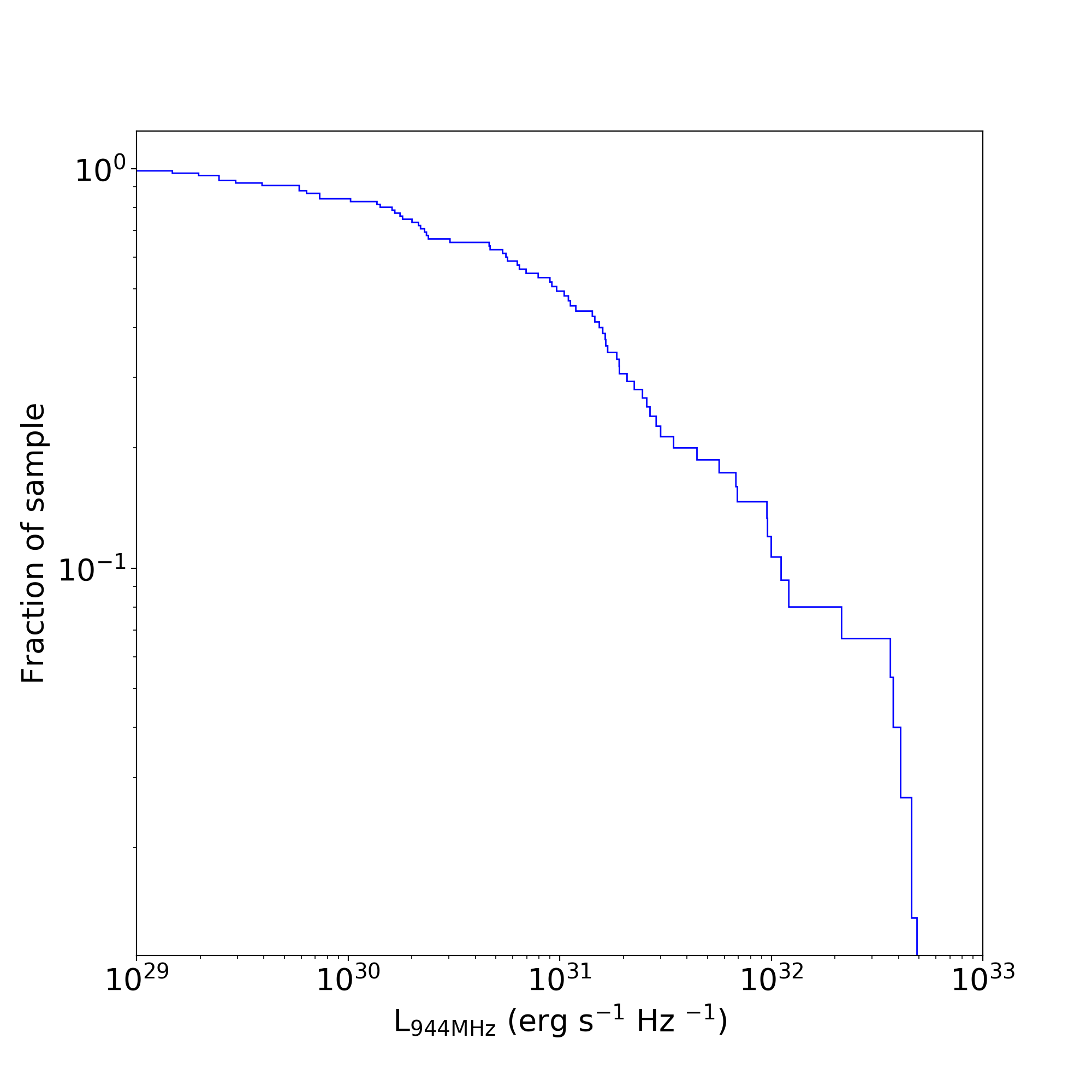}
   \includegraphics[width=9cm]{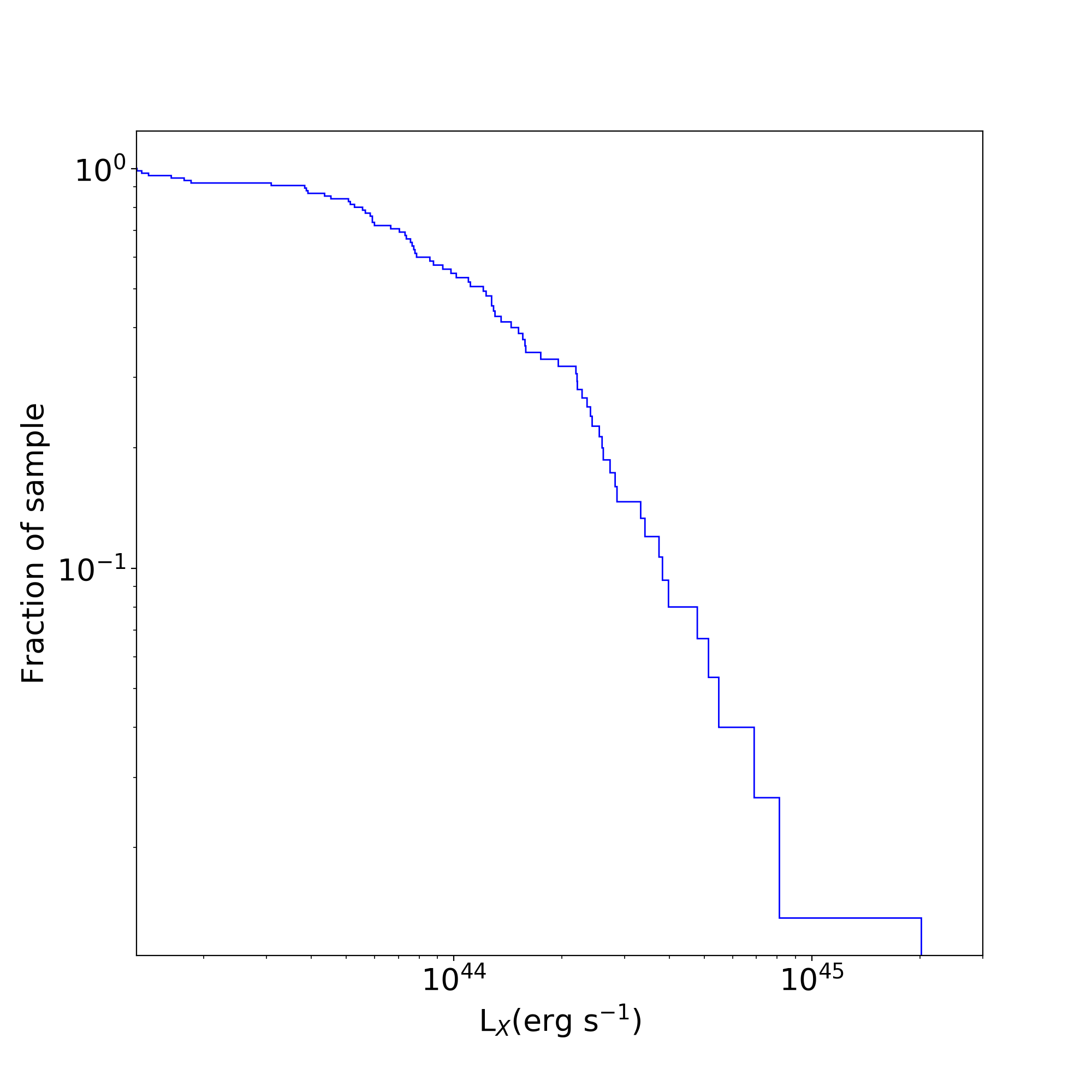}
   \caption{Radio and X-ray luminosities of the cluster sample. \emph{Left panel:\/} 944~MHz radio luminosity distribution for the cluster sample. \emph{Right panel:\/} X-ray luminosity distribution in the sample for the 0.5-2.0 keV band.}
         \label{N_Lum}
   \end{figure*}

In \Cref{N_Lum} we show the radio and X-ray luminosity distribution functions of the cluster sample at 944 MHz and 0.5 - 2.0 keV. The overall radio luminosities lie within a range of $\sim 10^{29}$ and $\sim 10^{33}$ erg s$^{-1}$ Hz$^{-1}$. The X-ray luminosities exhibit values from $\sim 10^{43}$ to $\sim 10^{45}$ erg s$^{-1}$.

%----------------------------------------------------------------- Flux Limit
%----------------------------------------------------------------- 

%----------------------------------------------------------------- Number Density BCG separation
%-----------------------------------------------------------------

%----------------------------------------------------------------- 
%----------------------------------------------------------------- 

\subsection{WISE colors}

\begin{figure*}
   \centering
   \includegraphics[width=9cm]{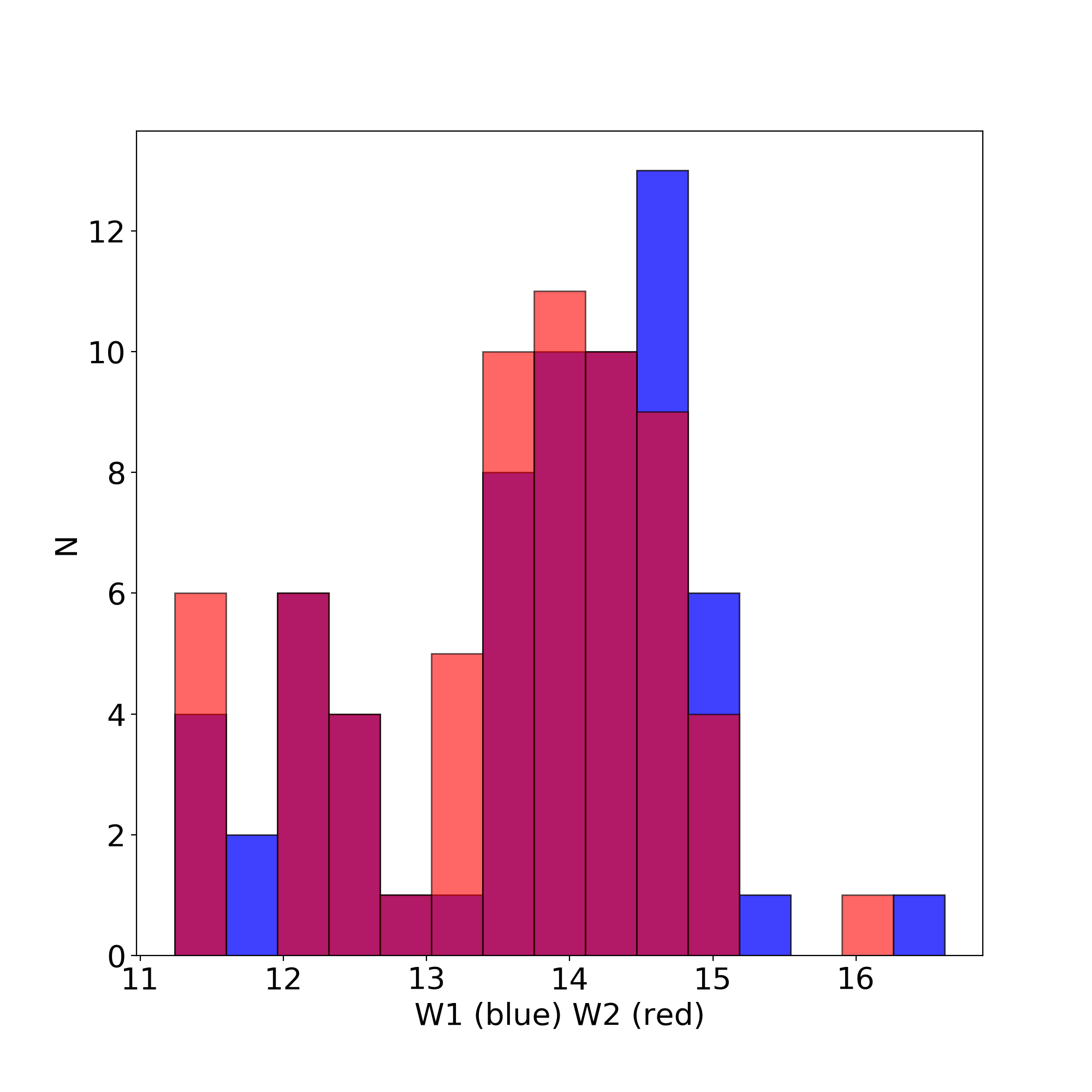}
   \includegraphics[width=9cm]{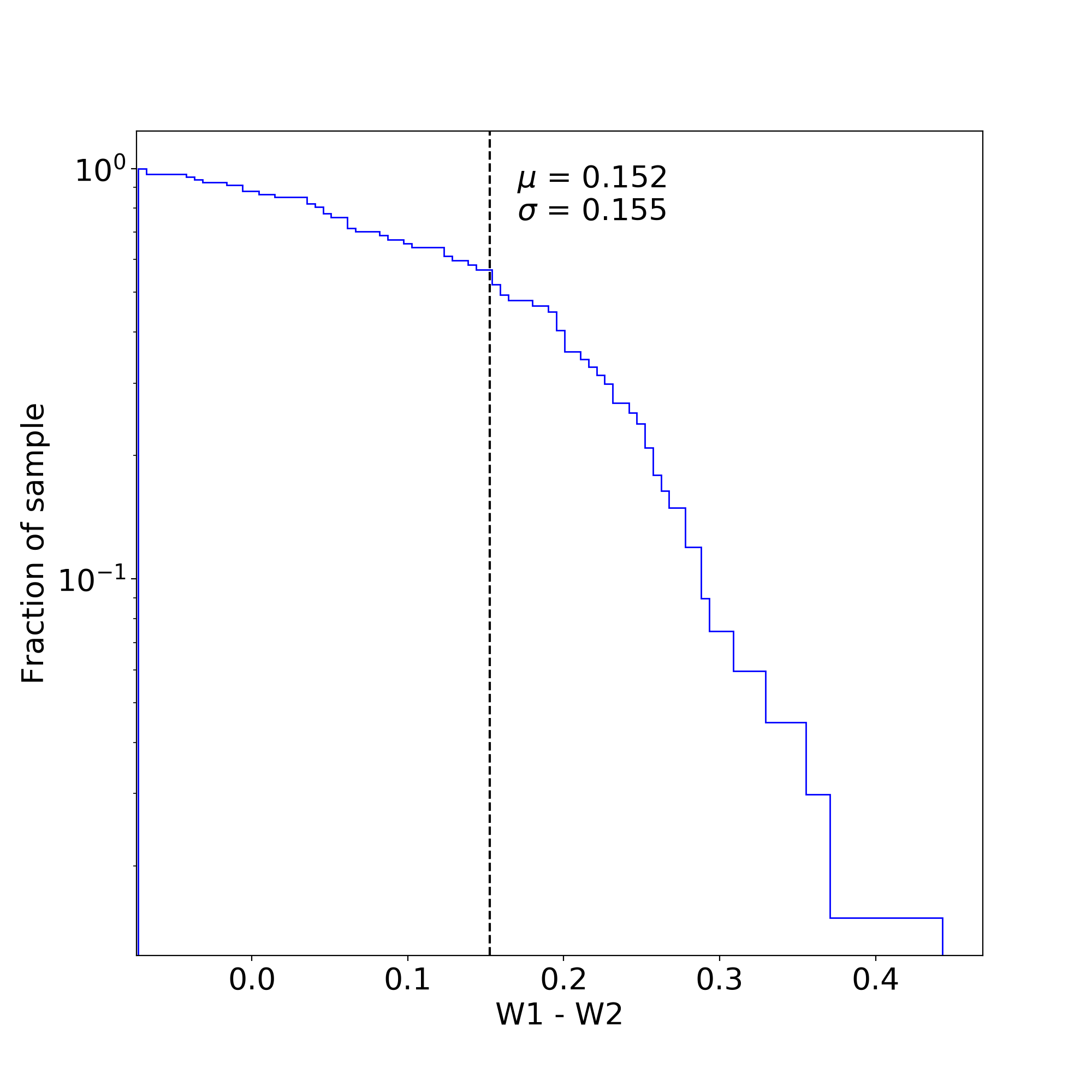}
   \caption{WISE colors of the sample. \emph{Left panel: \/} Histogram showing the magnitude distribution of WISE colors W1 (blue) and W2 (red) in Vega magnitudes. \emph{Right panel:\/} Distribution for W1 - W2.  }
         \label{W1_W2}
   \end{figure*}

One approach to identify AGN is a mid-infrared color criterion that is deduced from the separation between the power-law AGN spectrum and the blackbody stellar spectrum of galaxies, which has its peak at a rest-frame of 1.6 $\mu$m \citep{Assef_2010}. We applied this technique to our sample by using the Wide-field Infrared Survey Explorer (WISE) survey, which mapped the whole sky in four different bands: 3.4, 4.6, 12, and 22 $\mu$m, referred to as W1, W2, W3, and W4, respectively \citep{Wright_2010}. The color criterion we used is the difference of the magnitudes of W1-W2 (i.e., 3.4 - 4.6). For our sample, the W1 and W2 magnitudes and the W1-W2 criterion are shown in \Cref{W1_W2}. Our median value for the color criterion is $\mu \approx$ 0.152, with a corresponding interquartile range of $\sigma \approx$ 0.155. This is in contrast to \citet{Stern_2012}, who reported a value of W1-W2 $\geq$ 0.8 for their AGN selection, and \citet{Assef_2018}, who reported a value of W1-W2 $\geq$ 0.77. However, \citet{LaMassa_2019} and \citet{Mountrichas_2019} studied AGN in stripe 82 and showed that two-thirds of the X-ray detected AGN are not identified via the mid-infrared criterion. Especially AGN with luminosities between $10^{42.5} < L_X < 10^{44}$ erg s$^{-1}$ are not detectable by the WISE criterion and show bluer W1-W2 colors. In this population, the AGN does not seem to dominate the mid-infrared emission, and therefore, the color criterion is not applicable to our sample, where most of our objects meet this luminosity.

%One reason for this difference could arise from the different composition of their sample, as they are looking at AGN out to a redshift of z = 3.5 which is in contrast to our sample with a limit of z = 0.8. At very high redshifts even small amounts of dust extinction redden the W1-W2 colors. Another reason why the 0.8 cut should be handled with care is the fact that dilution by the host galaxy will cause bluer W1-W2 colors. This makes less powerful AGN non-detectable using the simple mid-infrared color criterion. 

%----------------------------------------------------------------- 
%----------------------------------------------------------------- 

%----------------------------------------------------------------- Analysis

\section{Analysis and discussion}
\label{Analysis_sec}

\subsection{Linear size of the BCGs}

Radio galaxies have a wide range of sizes and shapes, such as giant radio galaxies with largest linear sizes (LLS) of more than 0.7 Mpc, which were examined in \citet{Dabhade_2020}, for example, and small radio galaxies (e.g., \citet{Baldi_2015}). \citet{Hardcastle_2019} examined the relation between radio power and the linear size of a sample of 23344 radio-loud AGN, which is also referred to as the $P-D$ diagram (e.g., \citet{Turner_2017}). This diagram and the location of each source in it is an indicator for its initial conditions and its evolutionary state. The tracks of a source are associated with different phases in the evolution of the source. Objects with specific properties follow tracks on the plane that are mainly defined by the physics of the object. Remnant sources with switched-off jets describe a different set of tracks \citep{Hardcastle_2018}. However, the environment can also have an impact on the $P-D$ track. Sources in denser environments typically appear brighter than when they lie in more diluted environments \citep{Turner_2015, Yates-Jones_2022}. \\

\begin{figure}[h!]
   \centering
   \includegraphics[width=9cm]{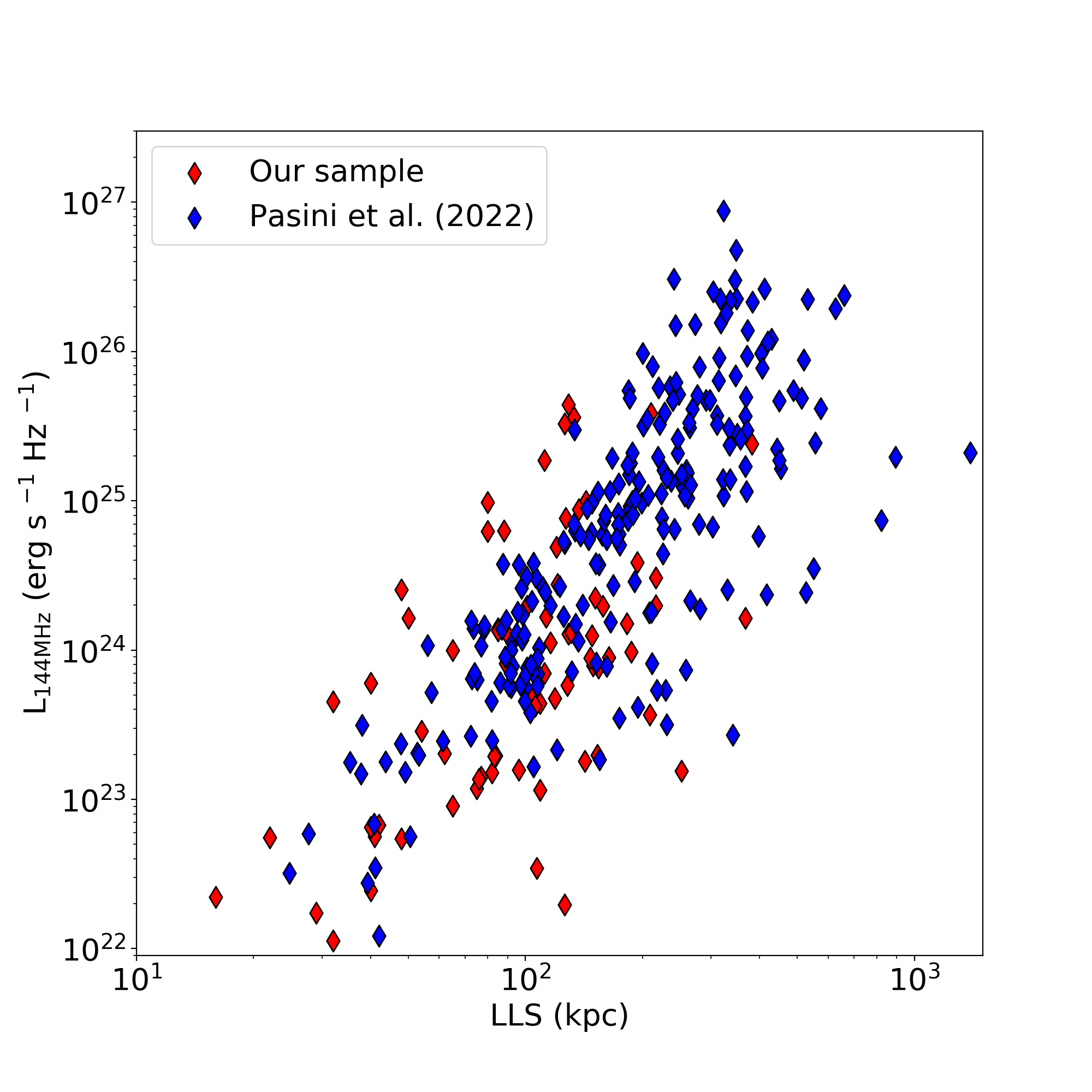}
   \caption{Projected LLS vs. the 144 MHz luminosity from P21 and our sample. We rescaled the luminosity from 944 MHz to 144 MHz using the spectral index $\alpha = 0.6$. }
         \label{P_D}
   \end{figure}

We measured the angular sizes of the sources and then converted them into linear sizes using their redshift. The errors on the angular sizes were taken to be equal to the synthesized beam. The $P-D$ diagram for our sample is shown in \Cref{P_D}. \citet{Pasini_2022} (P22 hereafter) examined a sample of 542 galaxy clusters and groups that were detected in the early performance verification phase (eFEDS) of eROSITA and compared them to the emission of the central radio galaxies detected with LOFAR. In \Cref{P_D} we compare the projected sizes of P22 to our data. To this end, we rescaled the 944 MHz luminosity to a luminosity at the LOFAR HBA central frequency of 144 MHz using a spectral index $\alpha = 0.6$. Our sample of eRASS:1 clusters extends the $P-D$ diagram to lower LLS values, with the luminosities being comparable to the luminosities $\mathrm{L}_{144\mathrm{MHz}}$ examined in P22, namely between $\sim 10^{23} $and$ \sim 10^{26}$erg s$^{-1}$Hz$^{-1}$. \\

The EMU sample reaches similar radio powers as the LOFAR data because EMU reaches a depth of 25–30 $\mu$Jy/beam at 944 MHz, which responds to a depth of 110 $\mu$Jy/beam at the LOFAR frequency of 144 MHz, while the LOFAR observation of the eFEDS field reaches $\sim$ 100 $\mu$Jy/beam. We also note that the resolution of ASKAP is $\sim$ 18 arcsec, in contrast to LOFAR, which reaches $\sim$ 6 arcsec \citep{Shimwell_2017, Shimwell_2019}.  \\

The interpretation of the $P-D$ diagram should be handled with care because several facts have to be taken into account. On the one hand, the environment and location within the cluster of the radio galaxy affects the position and track on the diagram. We also only observe the projected LLS, and we did not account for selection effects against large low-luminosity sources \citep{Shabala_2008, Turner_2015}. In addition, the redshift dependence might affect the $P-D$ diagram and its slope. We did not take the individual redshifts into account, which may result in a bias that distorts the true correlation between LLS and radio power. Different redshifts correspond to different cosmic epochs, in which the properties and evolutionary stages of radio galaxies can vary. Without a correction for redshift, objects at different redshifts may not be directly comparable.
However, the correlation between the LLS and the radio luminosity is clearly positive: larger radio galaxies usually have higher radio luminosities. \\

\begin{figure}[h!]
   \centering
   \includegraphics[width=9cm]{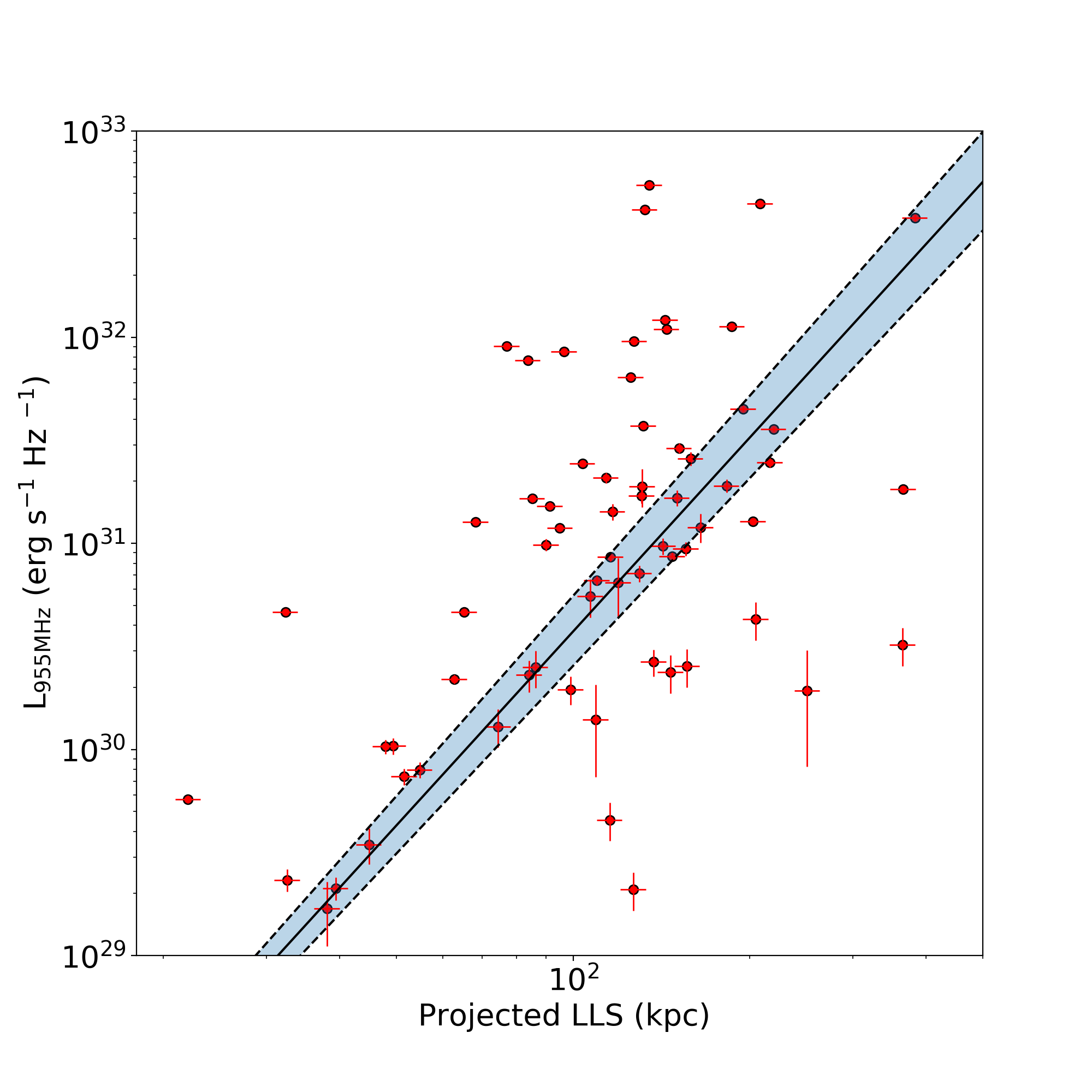}
   \includegraphics[width=9cm]{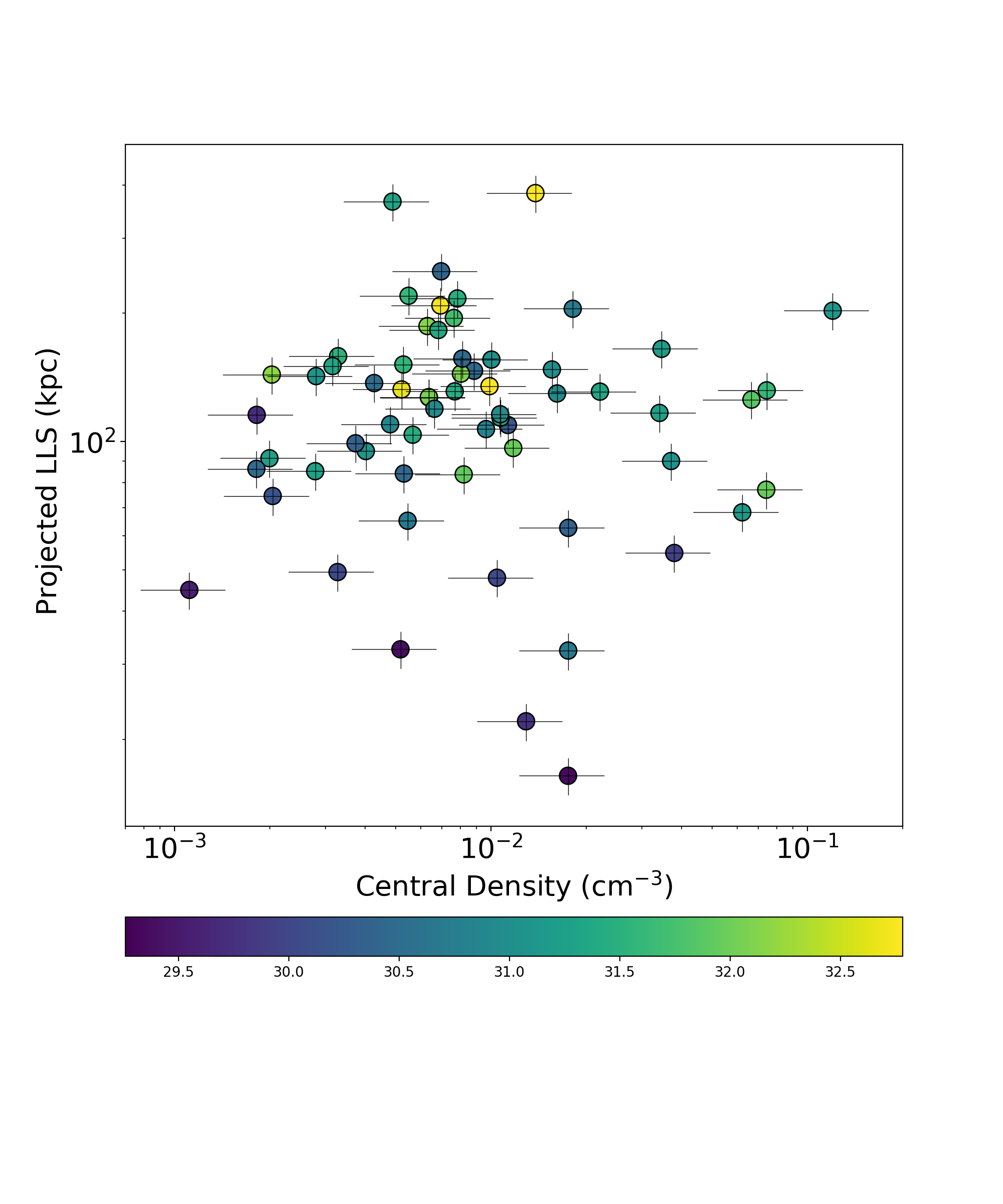}
   \caption{Projected LLS vs. radio power and central density. \emph{Upper panel:\/} Projected LLS vs. 944 MHz radio power of the EMU radio galaxies. The black line represents the best fit, and the blue region inside the dashed lines is the error band: $\log P_\mathrm{R} = (3.12 \pm 0.10) \cdot \log \mathrm{LLS} + (17.33 \pm 0.22)$. \emph{Lower panel:\/} Projected LLS of the radio source vs. the central density of the cluster. The data points are color-coded with the logarithmic radio luminosity in erg s$^{-1}$ Hz $^{-1}$. }
         \label{LLS_plots}
   \end{figure}

In \Cref{LLS_plots} we show the projected LLS versus 944 MHz radio power of the EMU radio galaxies. %We excluded unresolved sources that are smaller than the beam size. %
As previously observed, the correlation between LLS and luminosity is positive: larger radio galaxies are more powerful \citep{Owen_2002, Kolokythas_2018, Pasini_2022}. The mean value of the LLS is 130 kpc, with a standard deviation of 74 kpc. We calculated the relation between the radio power and the LLS to be $\log P_\mathrm{R} = (3.12 \pm 0.1) \cdot \log \mathrm{LLS} - (17.33 \pm 0.22)$. Nonetheless, we note that we also expect a diagonal sensitivity limit because for a given luminosity of an extended source at a given redshift, larger sources are harder to detect because their surface brightness is lower \citep{Shabala_2008}. For our sample, the theoretical cutoff limit does not play a role because it is four orders of magnitudes fainter that our measurements. \\

We also plot the projected LLS versus the central density of the ICM in \Cref{LLS_plots}. We conclude that these two observables are not correlated. However, color-coding the individual data points by the radio luminosity reveals that sources with a low central density and small LLS tend to exibit low radio luminosities. The lower central ICM density could imply a lower pressure in the radio lobes, leading to lower synchrotron emissivities, while smaller sources also favor lower luminosities.

%Both, the central density and the LLS determine the radio luminosity.

%----------------------------------------------------------------- Sample Comparison
%-----------------------------------------------------------------
%----------------------------------------------------------------- LLS Plots
%----------------------------------------------------------------- 

\subsection{BCG offset}

In order to examine the BCG offset of each cluster, we calculated the physical distance between the X-ray center given in the eRASS:1 cluster catalog and the optically identified BCG. The result is shown in the lower panel of \Cref{N_Offset}. The majority of BCGs are found within a radius of $\sim$ 100 kpc around the cluster center, which has been defined by the X-ray peak. This is consistent with the assumption of AGN feedback because the gas cooling out of the hot ICM can feed the central supermassive black hole (SMBH), while outer galaxies need to rely on more episodic triggers. \citet{Pasini_2021} conducted a phase-space analysis by comparing the clustercentric velocity with the clustercentric offset of the hosted galaxies to investigate the assembly and accretion history of these objects. Their analysis suggests that powerful radio galaxies are always located close to the cluster center. The interpretation was that the cooling ICM can feed the AGN if the galaxy lies close to the cluster density peak, where the cooling is more efficient. Nonetheless, galaxies located in cluster outskirts can also host radio AGN, and triggers such as mergers or interactions might be important \citep{Marshall_2018}. Small BCG offsets of less than 100 kpc are expected and found in most relaxed clusters because minor mergers can produce sloshing and displace the X-ray emission peak from the BCG \citep{Hamer_2016, Ubertosi_2021}. Larger offsets are usually an indication for major merger events, and therefore indicate strongly disturbed clusters \citep{2010A&A...513A..37H,Rosetti_2016, DePropris2021, Ota_2023, Seppi_2023}. In the next section, we compare the BCG offsets to the dynamical state of the clusters.

\subsection{Morphological parameters}

\begin{figure}
   \centering
   \includegraphics[width=9cm]{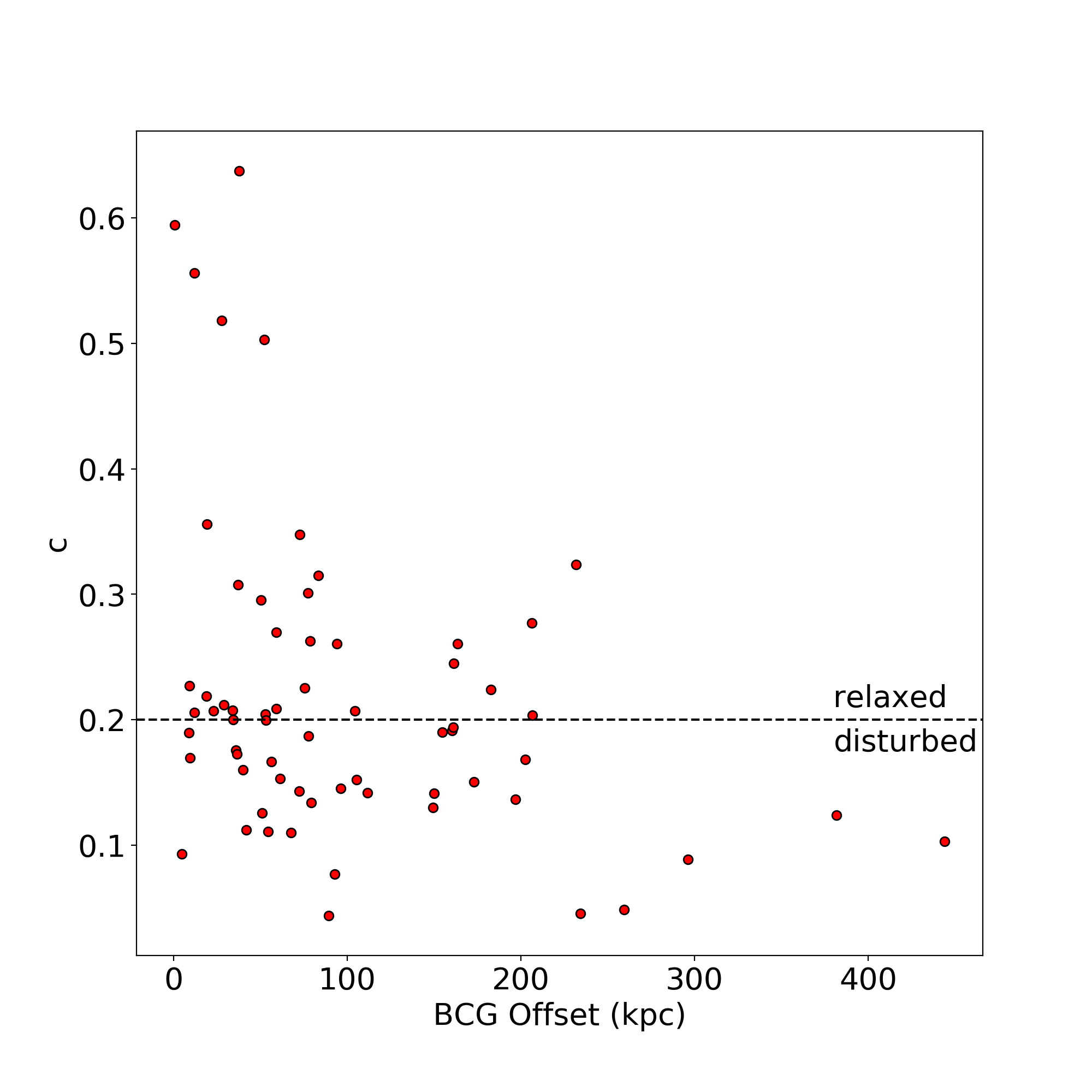}
   \caption{Concentration parameter plotted vs. the BCG offset. The median value of $c= 0.2$ subdivides the sample into relaxed and disturbed clusters.}
         \label{Morph_Param}
   \end{figure}

X-ray observations can be used to compute a quantitative measure of the dynamical status of a cluster.  Different morphological parameters for quantifying the dynamical status of a cluster have been described in literature. We focus on the concentration parameter $c$. The concentration parameter is the ratio of the X-ray flux within a radius of 100 kpc around the cluster center to the X-ray flux within a radius of 500 kpc \citep{Santos_2008}. It is defined as the ratio of the peak over the surface brightness $S$ as
\begin{align}
    c \equiv \frac{S(r <100\, \mathrm{kpc} )}{S(r<500\,\mathrm{kpc})}.
\end{align}
Clusters with a compact core that has not been disrupted by merger activity have higher concentration parameters. Hence, disturbed systems yield lower values for $c$. \citet{Cassano_2010} and \citet{Bonafede_2017} for example, have stated that considering the median value of $c= 0.2,$ it is possible to distinguish between disturbed ($c < 0.2$) and more relaxed ($c > 0.2$) clusters. \Cref{Morph_Param} shows the plot of the concentration parameter against the BCG offset. Clusters that show a large offset from the BCG to the X-ray center clearly have concentration parameters of $c < 0.2$ and can thus be classified as disturbed systems. The link between the dynamical status of the clusters to their BCG offsets is clear, and larger offsets are found in more strongly disturbed systems.\\

Another commonly used morphological parameter is the power ratio $P_3 / P_0$ (\cite{Buote_1995}). However, the number of photons in the eRASS:1 data is too low to yield a reliable estimate for the parameter $P_3 / P_0$. We therefore will return to this in future work.

%----------------------------------------------------------------- 

\subsection{Correlation of the radio and X-ray luminosity}

\begin{figure}
   \centering
   \includegraphics[width=9cm]{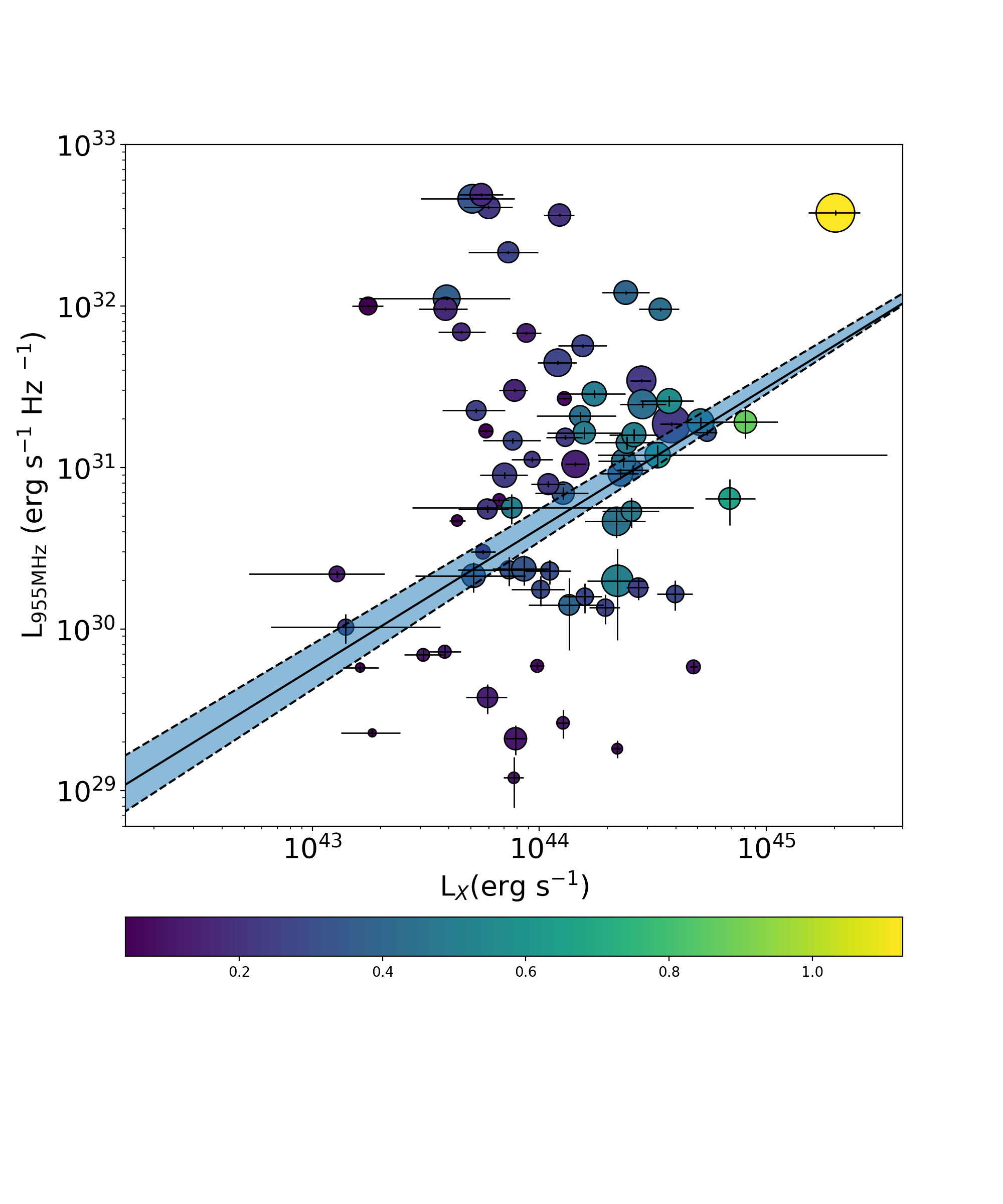}
   \caption{944 MHz luminosity of the central radio galaxy vs. the X-ray luminosity of the galaxy cluster in the 0.1-2.4 keV band. Each point is colored by redshift and scaled by the LLS. The black line represents the best fit, and the blue region inside the dashed lines shows the error band: $\log L_\mathrm{R} = (0.89 \pm 0.04) \cdot \log L_{\mathrm{X}} - (8.52 \pm 1.44)$.}
         \label{X_ray_Radio_Plots}
   \end{figure}

In this subsection, we investigate how the radio luminosity of the BCG relates to the global X-ray properties of the host cluster. \Cref{X_ray_Radio_Plots} shows the 944 MHz luminosity of the central radio galaxy versus the X-ray luminosity of the galaxy cluster in the 0.1-2.4 keV band. The colors display the redshift, and the size of the points the LLS. There is a trend for more luminous radio galaxies to be hosted in more X-ray luminous clusters, but the scatter is significant. Using the Python package \texttt{hyperfit}\footnote{https://github.com/CullanHowlett/HyperFit} , we calculated the correlation of the X-ray and radio luminosities \citep{hyperfit_2015}. This package provides a method for fitting a line  to data allowing for both intrinsic scatter and (potentially correlated) errors on all the input variables, here x and y for our 2D fit. We ran this program on our data with a fully converged MCMC (Markow-Chain-Monte-Carlo) run. We obtained $\log L_\mathrm{R} = (0.89 \pm 0.04) \cdot \log L_{\mathrm{X}} - (8.52 \pm 1.44)$. The p-value of this fit equals 0.05. We therefore consider this relation to be statistically significant.\\

Clusters that host radio sources with high radio luminosities ($> 5 \cdot 10^{31} \mathrm{erg s}^{-1} \mathrm{Hz}^{-1}$) broaden the function and introduce large scatter into the correlation. A similar correlation was found by \citet{Hogan_2015}, even though they did not quantify their results. 
Our results are also consistent with the best-fit relation found by \citet{Pasini_2022}, who find $\log L_\mathrm{R} = (0.84 \pm 0.09) \cdot \log L_{\mathrm{X}} - (6.46 \pm 4.07)$ and \citet{Pasini_2021, Pasini_2020}. We refer to \Cref{Correlation_Table} for the corresponding relations. Although the results agree, our sample clearly has less scatter in the relation. One reason for this could be the large number of radio upper limits that were taken into account in former work. which adds larger uncertainties. Generally, we would expect more scatter at lower frequencies as lower frequencies are emitted by electrons that are older. \\

%Hence we conclude: As galaxy clusters are extended nature the observed regions are of much greater physical extent than just the clusters central region. The cluster mass is therefore the main driver for the overall cluster integrated X-ray luminosity, with more massive clusters corresponding to more X-ray luminous systems. 

%The correlation that we observe is an indication that the most massive clusters host the most powerful radio-AGN.

%----------------------------------------------------------------- Correlation Plot
%----------------------------------------------------------------- 

%----------------------------------------------------------------- Correlation Table
\begin{table*}[!h]
%\begin{center}
\caption[]{Overview of the X-ray/radio correlation found by other authors.}
\label{Correlation_Table}
\begin{tabular}{lllll}
\hline
Author & Sources & $z$  & M & Correlation \\
\hline
\citep{Mittal_2009}& 64 & 0.0037 - 0.2153 & $10^{13}$ < M$_{\odot}$ < $10^{14}$ & $\log L_\mathrm{R} = (1.38 \pm 0.16) \cdot \log L_{\mathrm{X}} - (1.52 \pm 0.3)$ \\
\citep{Pasini_2020}& 247 &0.08 - 1.75 & $10^{13}$ < M$_{\odot}$ < $10^{14}$ & $\log L_\mathrm{R} = (1.07 \pm 0.12) \cdot \log L_{\mathrm{X}} - (15.90 \pm 5.13)$ \\
\citep{Pasini_2021}& 79 & 0.08 - 1.53 &  $8 \times 10^{12}$ < M$_{\odot}$ < $3 \times 10^{14}$ & $\log L_\mathrm{R} = (0.94 \pm 0.43) \cdot \log L_{\mathrm{X}} - (9.53 \pm 18.19)$ \\
\citep{Pasini_2022}& 542 & 0.1 - 1.3 & $3.4 \times 10^{12}$ < M$_{\odot}$ < $6.4 \times 10^{14}$ & $\log L_\mathrm{R} = (0.84 \pm 0.09) \cdot \log L_{\mathrm{X}} - (6.46 \pm 4.07)$ \\
This work & 75 & $0.03 - 1.1$ & $10^{13}$ < M$_{\odot}$ < $10^{15}$ & $\log L_\mathrm{R} = (0.89 \pm 0.04) \cdot \log L_{\mathrm{X}} - (8.52 \pm 1.44)$ \\
\hline
\end{tabular}
\label{specs}
%\medskip\\
%\end{center}
\end{table*}
%----------------------------------------------------------------- 

\subsection{Mechanical powers of the jet}

The radio lobes only radiate a small fraction of the total power  away that is supplied to the lobes when the source is active. This small fraction of radiation is the radio luminosity, which is only a fraction of the energy produced by the AGN through accretion of matter toward the black hole itself. A larger fraction of the power is stored in the radio lobes and dissipated during the expansion of the jets into the ICM \citep{Smolcic_2017}.
A direct approach for calculating the mechanical power of a radio jet would be to derive it from the properties of the radio source in comparison with a radio source evolution model. In most cases, this is not possible because the radio environment of the sources is unknown. Moreover, the luminosity evolves with the age of the sources \citep{Turner_2015, Hardcastle_2018, Yates-Jones_2022}. A common approach to overcome this problem is to estimate the mechanical energy of the jet by estimating it directly from the radio luminosity \citep{Sabater_2019}. 

This conversion of the  mechanical power of the jet into radio luminosity is usually estimated from the cavities that are inflated by radio sources in the surrounding ICM as observed in X-ray images. The total mechanical energy is then calculated to $4pV$, with $p$ the pressure of the surrounding medium, and $V$ the volume of the cavity. The factor of 4 arises if the enthalpy of the relativistic plasma in the radio lobes, which is $3pV$ , is added to the work performed to inflate the cavities, which is $pV$. When an estimate of the source age (e.g., via the buoyancy timescale of the cavity) is also given, a lower limit of the mechanical power of the jet can be estimated. This is found to correlate with the observed radio luminosity \citep{Rafferty_2006, Birzan_2008, Cavagnolo_2010}. \\

\begin{figure*}[h]
   \centering
   \includegraphics[width=8.5cm]{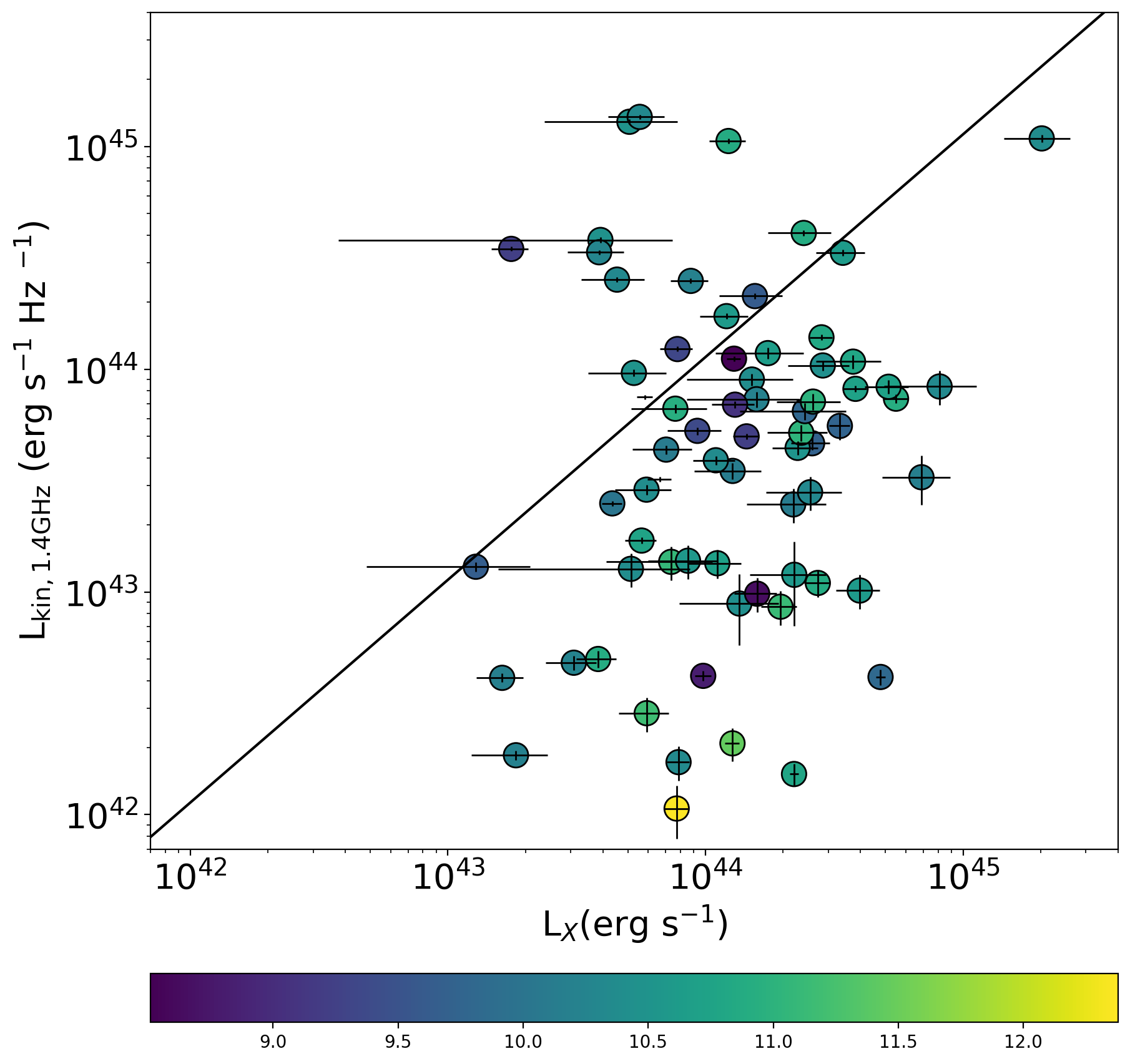}
   \includegraphics[width=9cm]{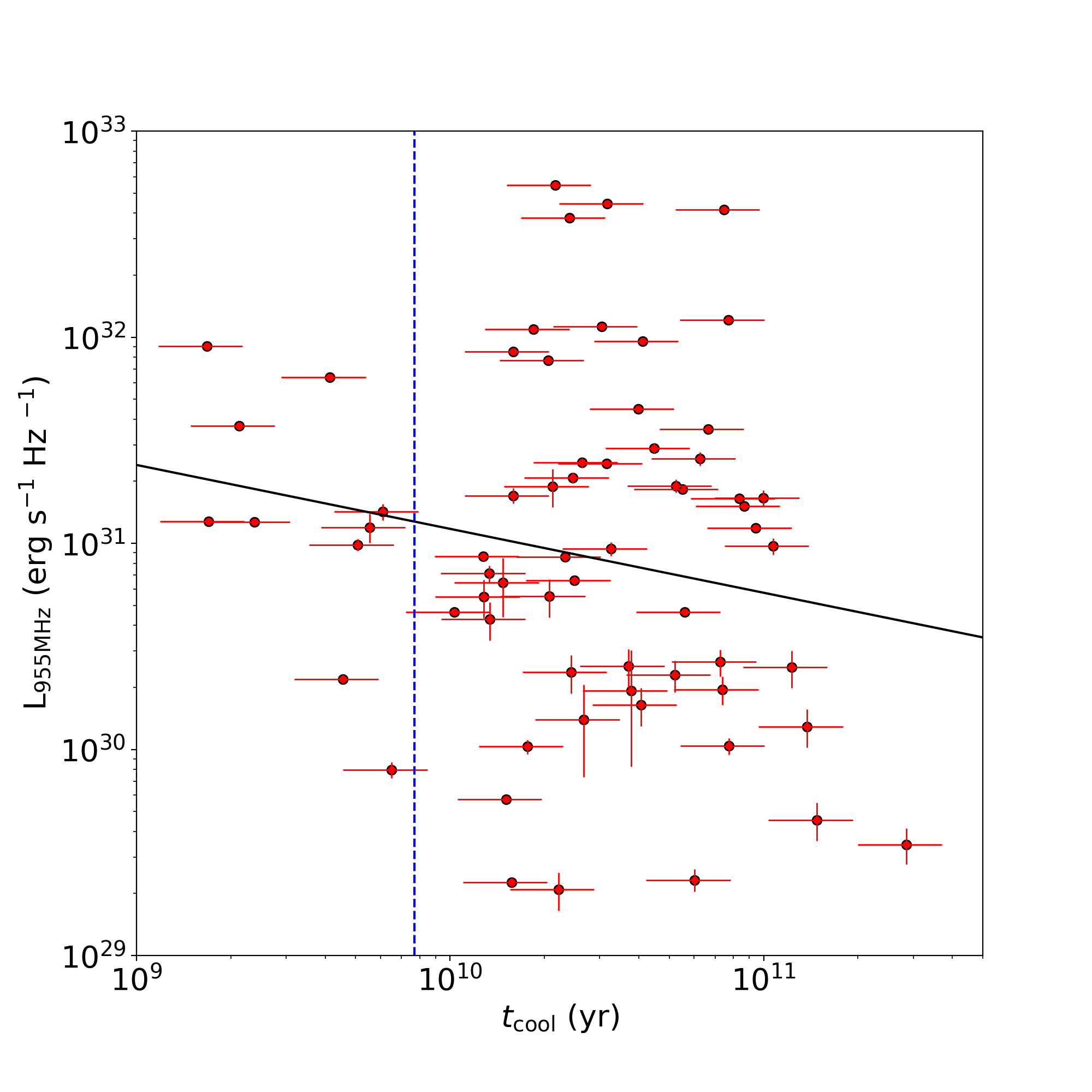}
   \caption{X-ray luminosity of the clusters sample vs. the corresponding radio luminosity of the central radio source and the 944 MHz radio luminosity of the central radio source vs. the central cooling time of each cluster. \emph{Left panel:\/}1.4 GHz kinetic luminosity extrapolated from 944MHz vs. the X-ray luminosity of the corresponding cluster. The individual points are color-coded with respect to the logarithmic central cooling time. \emph{Right panel:\/}  944 MHz radio luminosity of the central radio source vs. the central cooling time of each cluster. The vertical dashed blue line indicates $t_{\mathrm{cool}} = 7.7$ Gyr and separates CC and NCC clusters. The black line indicates the fit: $\log L_\mathrm{R} = (-0.31 \pm 0.03) \cdot \log \mathrm{t_{cool}} + (34.17 \pm 0.31)$}
         \label{L_kin}
   \end{figure*}

Another approach to relate the mechanical power of the jet to the radio luminosity is based on the synchrotron properties, and therefore, on the composition of the jet plasma ( \cite{Willot_1999}). \citet{Heckman_2014} found that both approaches provide consistent estimates of the mechanical powers of the jet and proposed a population-averaged conversion,

\begin{align*}
P_{\mathrm{mech,cav}} = 2.8 \times 10^{37} \bigg(\frac{L_{\mathrm{1.4 GHz}}}{10^{25}\mathrm{W Hz}^{-1}}\bigg)^{0.68}\mathrm{W}.
\end{align*}
Furthermore, the kinetic luminosity at a rest-frame frequency of 1.4 GHz is described by 
\begin{align*}
\log L_{\mathrm{kin, 1.4 GHz}} = 0.86 \log L_{\mathrm{1.4 GHz}} + 14.08 + 1.5 \log f_W.
\end{align*}
$L_{\mathrm{kin, 1.4 GHz}}$ describes the kinetic luminosity, and $L_{\mathrm{1.4 GHz}}$ is the luminosity at 1.4 GHz. $f_W$ is an uncertainty parameter, which is estimated to be around 15 from observations \citep{Smolcic_2017}. In order to determine the kinetic luminosity for our sample, we converted the radio power at 944 MHz to radio powers at a frequency of 1.4 GHz, assuming a spectral index of $\alpha = 0.6$. We then compared our results to the X-ray luminosity of the host cluster. We again used the package \verb|hyperfit| to estimate our relation in log-log scale in the form

\begin{align*}
Y = \alpha + \beta X + \varepsilon   ,
\end{align*}
 where $\alpha$ and $\beta$ represent the intercept and slope, and $\varepsilon$ is the intrinsic scatter. We find $\alpha = -3.81\pm 1.01, \beta = 1.08\pm0.03, $ and $\varepsilon = 0.91\pm0.61$ (see \Cref{L_kin}). This approximately agrees with the values found by \citet{Pasini_2022}, who found $\alpha = -2.19\pm 4.05, \beta = 1.07\pm0.11, $ and $\varepsilon = 0.25\pm 0.05$. However, we obtain a larger $\varepsilon$-factor for our sample, which represents a higher uncertainty in the observed values. As the conversion from the 944 MHz luminosity into kinetic luminosity at 1.4 GHz depends on a number of assumptions, we introduced large errors that result in a high scatter of the relation. Moreover, our sample is smaller than that of P22, which also results in a higher scatter. We also note that considering a sample over a wide redshift range can introduce a bias into this estimation \citep{Godfrey_2016}. Previous results from P22 stating that in most clusters the heating from the central AGN  balances the ICM radiative losses cannot be confirmed from our data (see also, e.g., \citet{McNamara_2012, McNamara_2016} for a review). However, we note that P22 used additional COSMOS data of lower-luminosity galaxy groups than the initial eFEDS data (see their Fig. 10). When the COSMOS data are removed from their data, their correlation between the kinetic luminosity at 1.4 GHz and $L_{\mathrm{X}}$ also becomes far lower. The scatter in the radio luminosity appears to increase strongly with $L_X$, similar to the results of \citet{Main_2017}. At values $L_X > 10^{43}$ erg s$^{-1}$ , the correlation disappears. The kinetic luminosity acts as a proxy for the heating rate, and the X-ray luminosity acts as a proxy for the cooling rate. Hence, the central AGN appear to counterbalance radiative losses from the ICM in low-luminosity clusters and groups, but this relation breaks for high-luminosity clusters. \citet{Main_2017} also investigated this relation and reported that the correlation between kinetic luminosity and X-ray observables only holds in clusters with short (< 1 Gyr) central cooling times. We derived the central cooling time $t_{\mathrm{cool}}$ based on the X-ray temperature and color-code the individual points with respect to $t_{\mathrm{cool}}$ in \Cref{L_kin} to examine whether $t_{\mathrm{cool}}$ affects the individual cluster position in the $L_{\mathrm{kin}} - L_{\mathrm{X}}$ diagram. We cannot confirm the results from \citet{Main_2017} based on our data. In high-luminosity clusters, the variability among the AGN population seems to be higher, resulting in a higher scatter in the $L_{\mathrm{kin}} - L_{\mathrm{X}}$ correlation. However, the measurements are only a snapshot in the lifetime of an AGN at a certain point in their duty cycle. Averaged over a longer period, AGN heating could still balance cooling, but the implication is that at higher $L_X$ , the AGN are more variable.

 %Still the trend suggests that heating from the central AGN may counterbalance radiative losses from the ICM in most clusters  \citep{Main_2017}. 
%----------------------------------------------------------------- Flux Limit
%----------------------------------------------------------------- 

\subsection{Cooling time}

Because AGN feedback heats the ICM and regulates its cooling rate, the correlation between central cooling time and radio luminosity can be used to investigate the relation between ICM cooling and AGN heating. 
%There are several tools for measuring the cooling properties of the hot ICM in the core of the cluster: Such as central entropy, classical cooling rate or the temperature. 
Because the eRASS:1 cluster catalog provides the central density $n_e$ and the temperature $T_X$ for all clusters, the central cooling time can be approximated as \citep{Sarazin_1986}

\begin{align*}
    t_{\mathrm{cool}} = 8.5 \cdot 10^{10} \mathrm{yr} \bigg( \frac{n_P}{10^{-3} \mathrm{cm}^{-3}} \bigg)^{-1} \bigg( \frac{T_{\mathrm{g}}}{10^8 \mathrm{K}}\bigg)^{\frac{1}{2}}.
\end{align*}
We assumed a hydrogen density $n_P = 0.83 n_e$ \citep{McDonald_2018}. We plot the cooling time $t_{\mathrm{cool}}$ versus the radio luminosity $L_{\mathrm{R}}$ in the right panel of \Cref{L_kin}. We removed two outliers with derived central cooling times of > $10^{12}$ yr because these values are most likely the result of incorrect densities and temperatures in the catalog data. While the plot is messy, there seems to be a trend of an anticorrelation between the two quantities. To quantify this anticorrelation, we find $\log L_\mathrm{R} = (-0.31 \pm 0.03) \cdot \log \mathrm{t_{cool}} + (34.17 \pm 0.31)$. We also indicate $t_{\mathrm{cool}} = 7.7$ Gyr because this value is commonly used to distinguish between cool-core (CC) and non-cool-core clusters (NCC). Our sample contains 10 CC clusters, while the rest are NCC clusters. \citet{Mittal_2009} examined a sample of 64 HIFLUGCS clusters and their central radio galaxies and found a similar trend for an anticorrelation between the cooling time of the cluster and the radio luminosity of its central AGN. For their sample, they reported a slope of $-3.16 \pm 0.38,$ which is in contrast to our relation, where we find a slope of $-0.31 \pm 0.03$. We note that \citet{Mittal_2009} used a central definition of 0.4$\%$ of r$_{500}$ , which is a radius that cannot be resolved by eROSITA. Hence, any comparison should be handled with care. 
%only examine clusters with cooling times of 8 Gyr < $t_{\mathrm{cool}}$ < 9 Gyr\commentTR{This is not correct otherwise Mittal+ would only have NCCs.} our findings extend their sample with CC and NCC clusters which exhibit cooling times of 9 Gyr < $t_{\mathrm{cool}}$ < 12 Gyr. Comparing the results indicate a flattening of the relation at longer cooling times of $t_{\mathrm{cool}}$ > 9 Gyr. Therefore we conclude that the trend of anti-correlation seems to be valid for both CC and NCC clusters but is 10 times stronger %in CC clusters. 
In general, this apparent anticorrelation could be indicative of a need for more powerful AGN in clusters with short central cooling times. Cool-core clusters with very short cooling times seem to need much more powerful AGN, unlike NCC clusters, in which this trend is less obvious. Finally, the cluster mass also appears to play a role. \citet{Bharadwaj_2014} investigated the same relation for galaxy groups and found no relation between the central cooling time and the radio luminosity of the central AGN. For a discussion of the difference of AGN feedback in clusters and groups, we refer to \cite{Pasini_2021}.

\subsection{Density profiles}

\begin{figure*}[h]
   \centering
   \includegraphics[width=9cm]{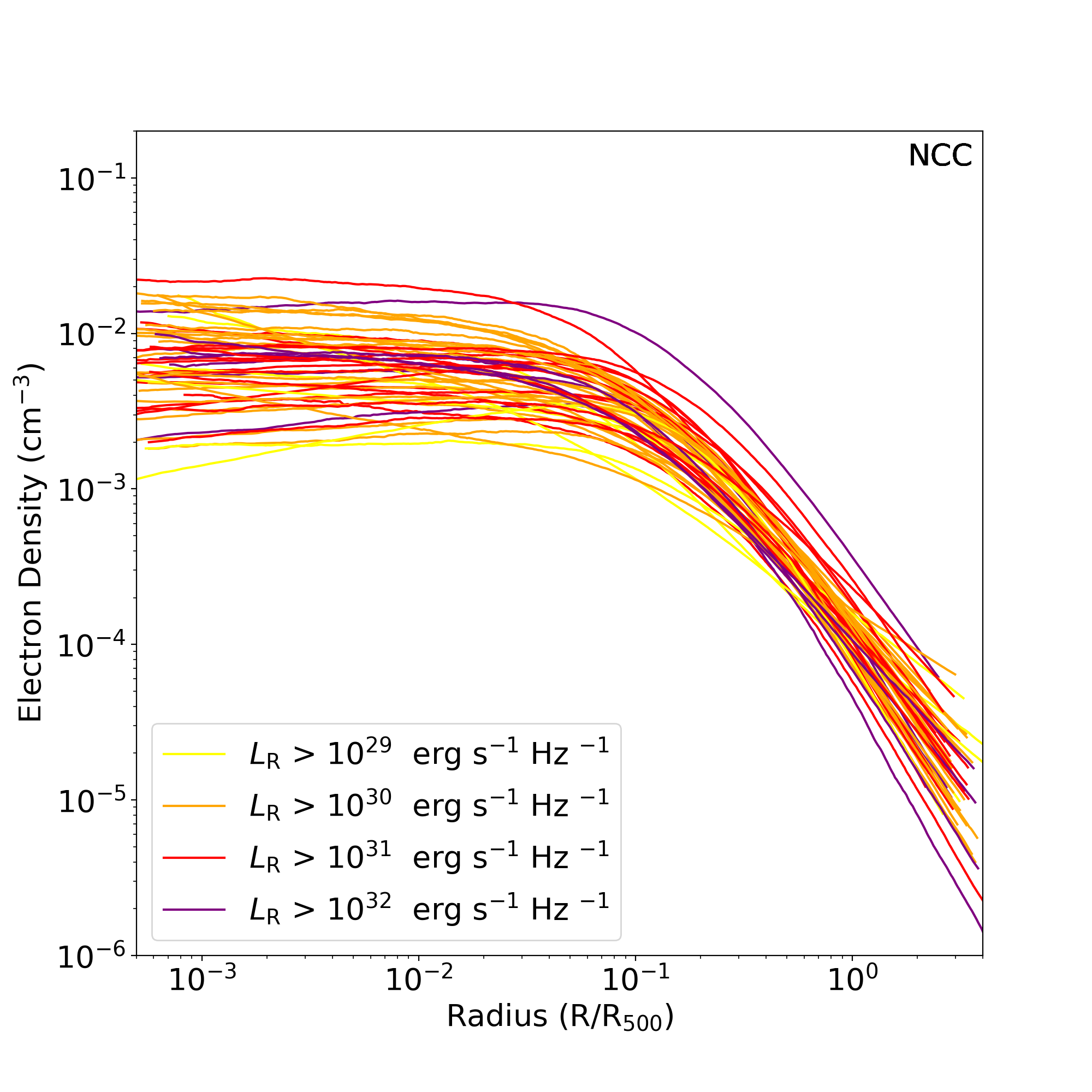}
   \includegraphics[width=9cm]{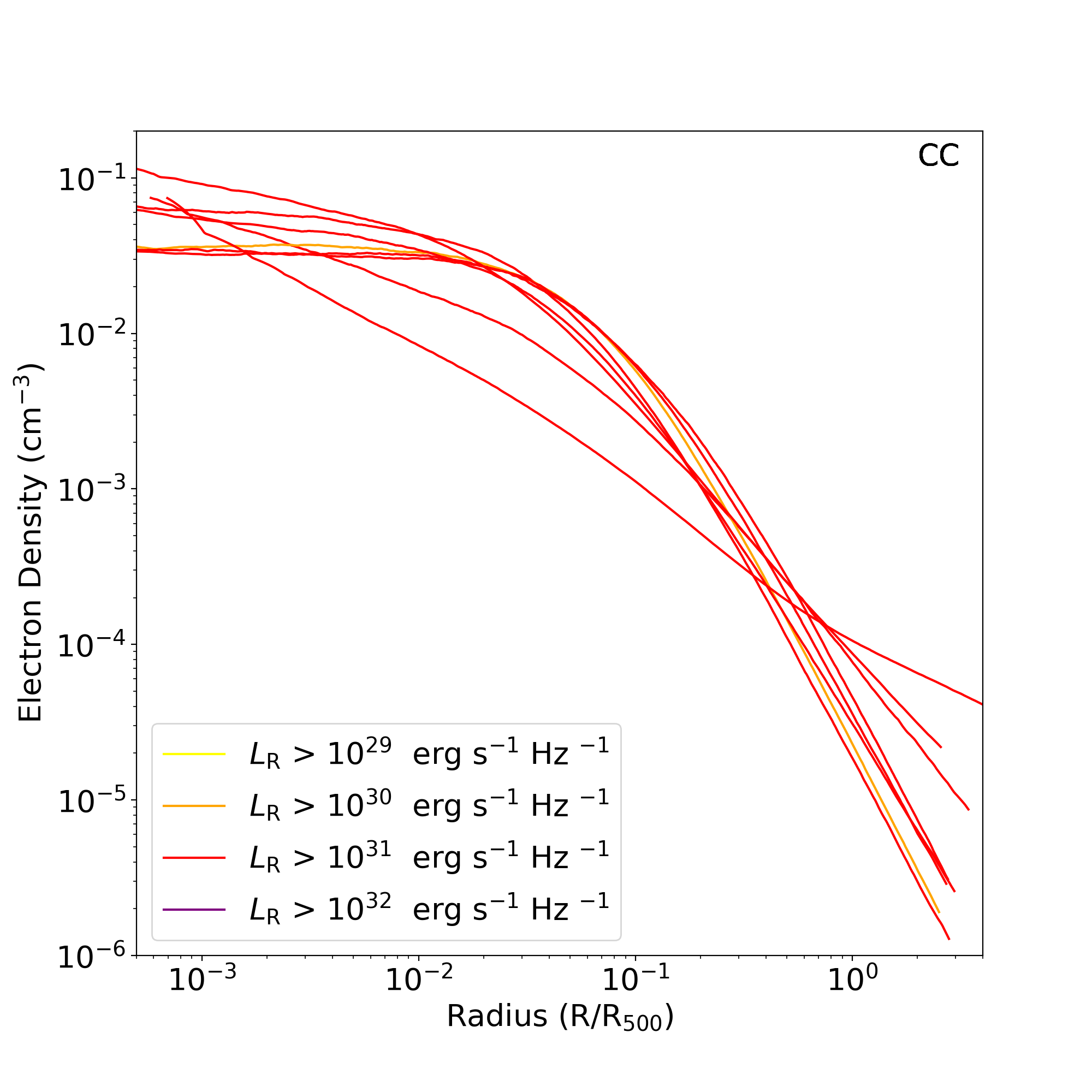}
   \caption{Electron density profiles of the cluster sample. \emph{Left panel:\/} Electron density profiles of all non-cool-core clusters vs. the radius scaled to R$_{500}$. The colors of the profiles represent the radio luminosity of the central radio source of the corresponding cluster. \emph{Right panel:\/}  Corresponding plot for the cool-core clusters.}
         \label{Density_profiles}
   \end{figure*}

In order to investigate the connection between CC and NCC clusters and their corresponding radio luminosity, we plot the density profiles of all clusters of our sample and color-code them by their radio power. The resulting plot is shown in \cref{Density_profiles}. We again removed the density profiles of two outliers with very low central densities of $n_e < 10^{-4}$ because we assume that these values are incorrect. Clusters with a higher central density ($n_e > 10^{-2}$cm$^{-3}$), subsequently CC clusters, tend to host more luminous radio sources with radio luminosities of $L_\mathrm{R} > 10^{31}$ erg s$^{-1}$ Hz$^{-1}$ . This confiRMS that CC clusters always host a powerful radio-mode AGN. Clusters with lower central densities ($n_e < 10^{-2}$cm$^{-3}$) show no connection, and we draw the conclusion that  NCC clusters host low- as well as high-luminous radio sources. This agrees overall with the general findings that X-ray cavities and therefore powerful radio-mode AGN are usually found within CC clusters (e.g.,  \citep{OSullivan_2011, Hlavacek-Larrondo_2012, Birzan_2020, Olivares_2022}). In contrast to this strong connection of AGN activity and CC clusters, NCC clusters and AGN activity do not appear to be correlated. \citet{Mittal_2009} showed that NCC clusters may also host strong radio AGN, which can be explained by merging activities or other mechanisms, for instance. 

%----------------------------------------------------------------- Density Profiles
%-----------------------------------------------------------------

\subsection{Noteworthy clusters}

\begin{figure*}[h]
   \centering
   \includegraphics[width=9cm]{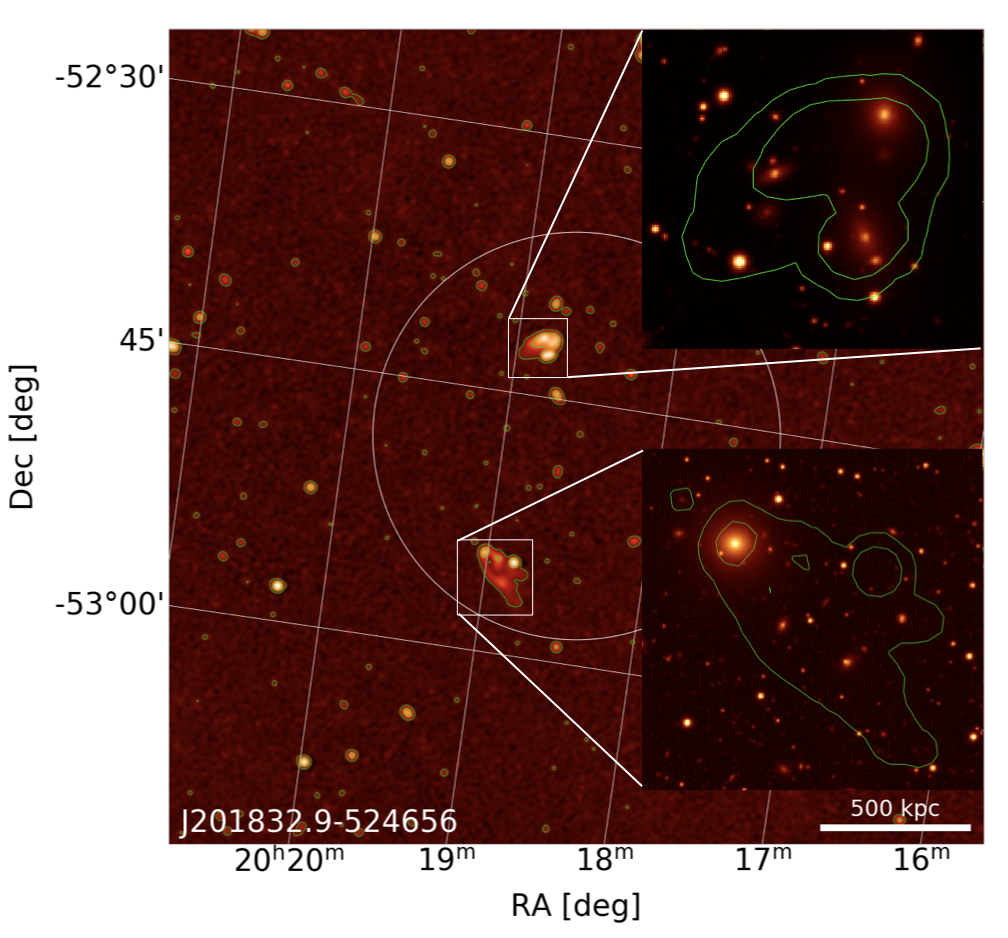}
   \includegraphics[width=9.25cm]{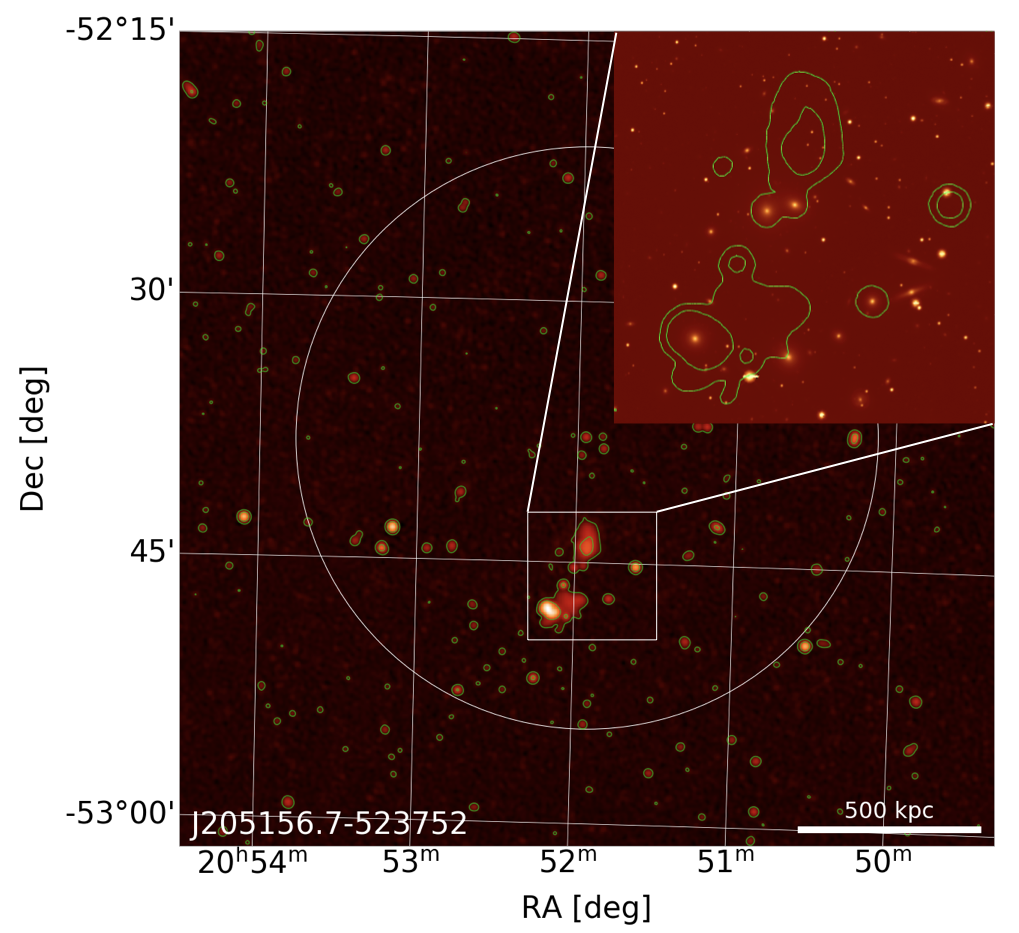}
   \includegraphics[width=9cm]{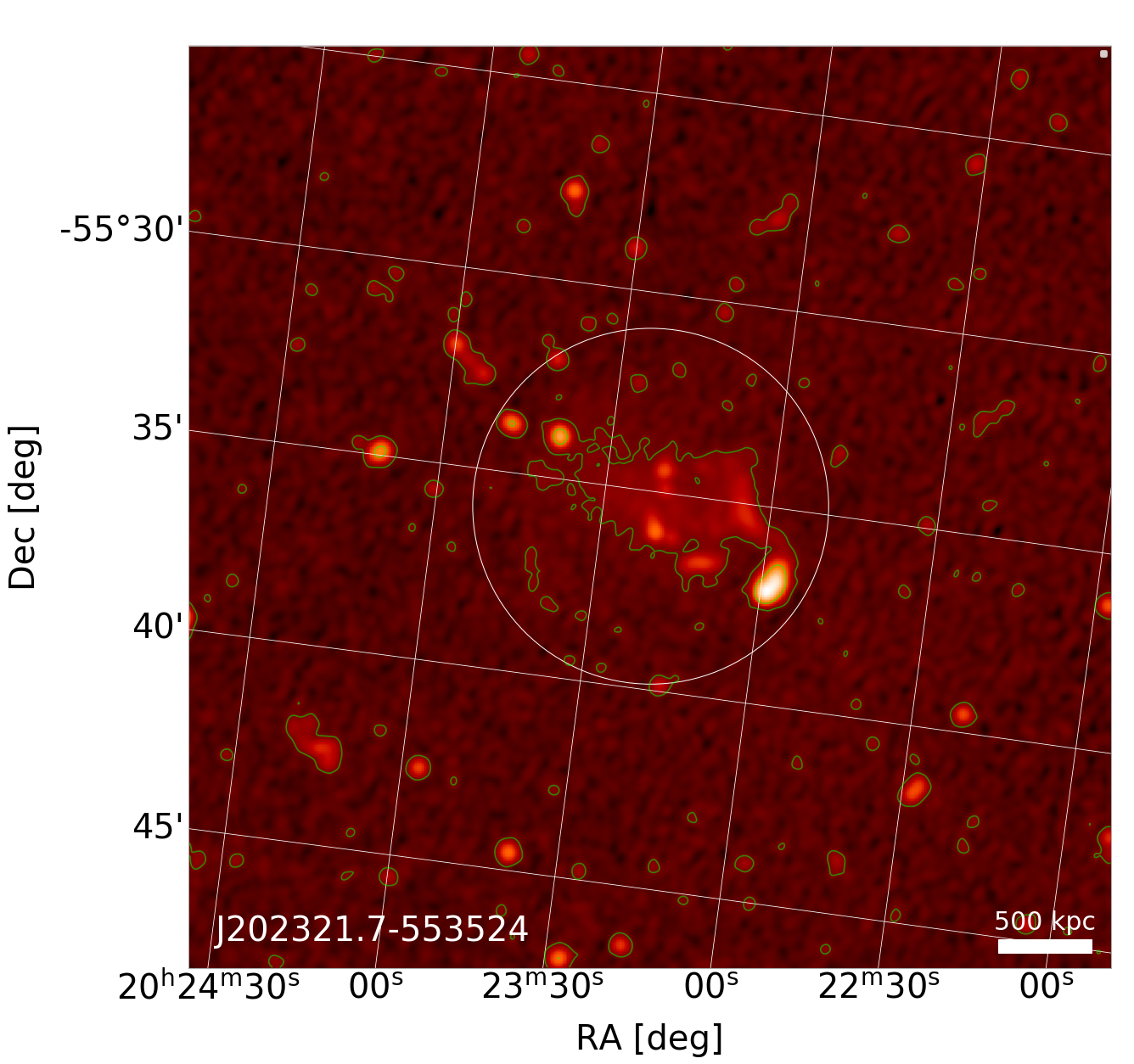}
   \includegraphics[width=9.1cm]{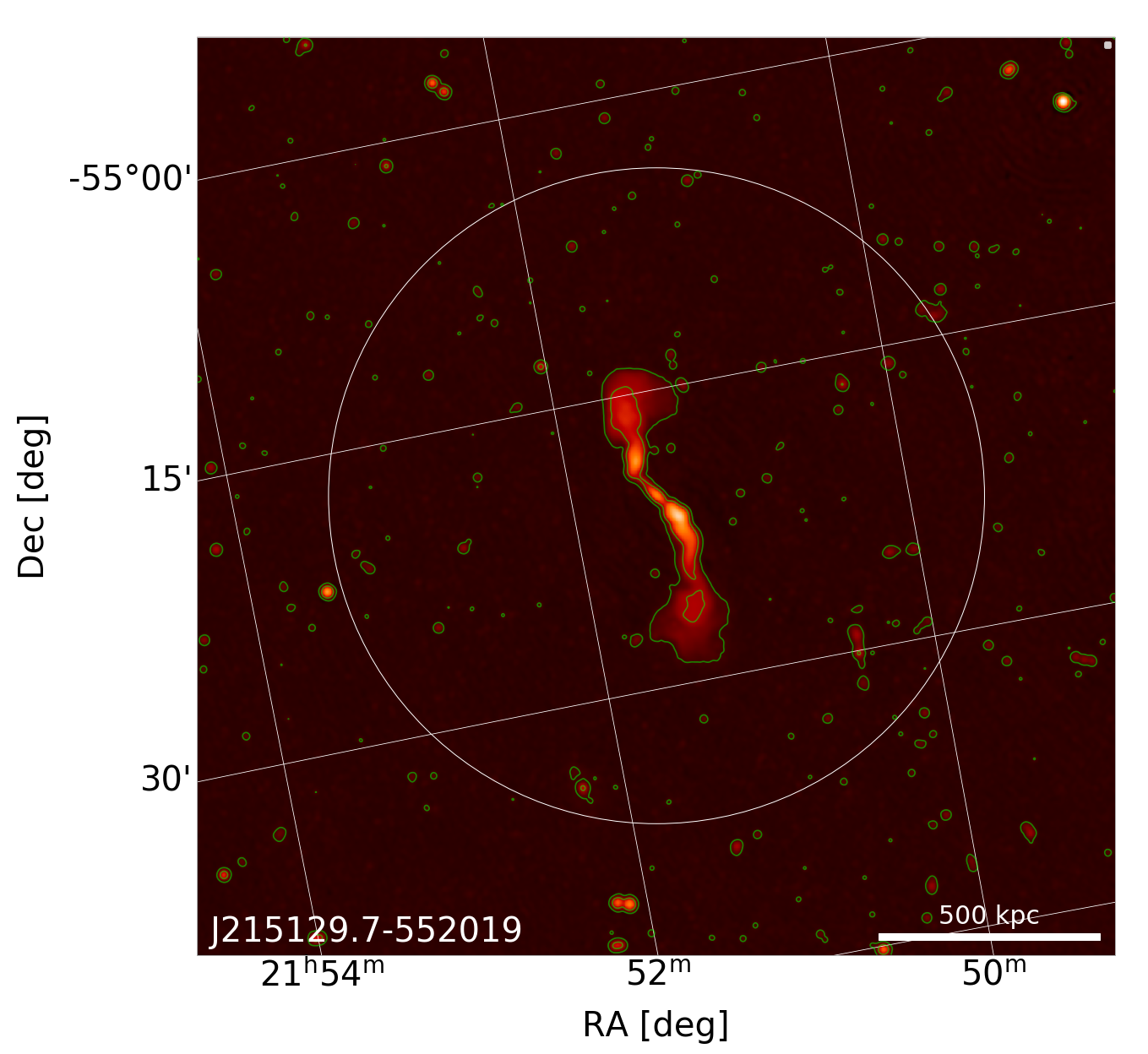}
   \caption{Special radio sources contained in the EMU field. \emph{ Upper left panel:\/} Radio cutout from the EMU image of cluster J201832.9-524656 (Abell S0861) at $z=0.05$ showing two elongated radio sources. The optical overlay reveals a complex interplay between at least three radio galaxies on the northern radio source and an infalling radio galaxy in the southern source. \emph{Upper right panel:\/} The radio cutout of J205156.7-523752 (PLCKESZ G345.4-39) at $z=0.04$ shows two connected elongated radio sources that seem to be hosted by at least three radio galaxies. \emph{Lower left panel:\/}  Radio image of J202321.7-553524 (SPT-CL J2023-5535) at $z=0.22$ revealing a large radio source. \emph{Lower right panel:\/} Cutout of the nearby cluster J215129.7-552019 (RXC J2151.3-5521) at $z=0.03$ clearly showing the radio jets that originate from the central radio source. All optical overlays made use of legacy survey DS9 data, and the white circle represents $R_{500}$.}
         \label{Special_Sources_1}
   \end{figure*}

Our cluster sample contains some interesting radio sources, four of which are presented in \Cref{Special_Sources_1}. The upper left panel shows galaxy cluster J201832.9-524656 (Abell S0861) at $z=0.05$ . The white circle is $R_{500}$. This cluster contains two interesting radio sources with very elongated shapes. We also show the overlay with optical data from the legacy survey DR9, in which the upper radio source presumably consists of at least three galaxies exhibiting radio emission. The shape of the radio emission suggests a complex interplay between these galaxies. The southern radio source in J201832.9-524656 resembles the shape of a jellyfish galaxy. The optical overlay shows a bright galaxy in the upper part of the radio galaxy. This shape suggests that this galaxy is moving northwest.\\

The upper right panel of \Cref{Special_Sources_1} shows cluster J205156.7-523752 (PLCKESZ G345.4-39) at $z=0.04$. This cluster hosts two elongated radio sources in the southern part of the cluster that seem to be connected. The optical overlay with legacy survey DR9 data reveals that the upper radio source consists of two near galaxies, and the upper radio source is at least one radio galaxy. The shapes of these sources suggest that these galaxies have either undergone some merging activities in the past or will do so in the future. In the lower left panel, we display cluster J202321.7-553524 (SPT-CL J2023-5535) at $z=0.22$. The radio image of this cluster reveals a radio source that covers large areas of the whole cluster, with a strong peak in the southeast. This radio source could be an indication for a radio halo and has also been studied by \citet{HyeongHan_2020}. In the lower right panel, we display the nearby cluster J215129.7-552019 (RXC J2151.3-5521) at $z=0.03$. The radio jets clearly originate from the central radio source in the cluster.
%----------------------------------------------------------------- Special Sources
%-----------------------------------------------------------------

\section{Conclusions}
\label{Conclusion_sec}

We used the eROSITA eRASS:1 cluster catalog and the ASKAP pilot survey EMU to examine the central radio galaxies hosted in galaxy clusters covered in the EMU survey. We can draw the following conclusions:

   \begin{itemize}
      \item[] Our sample consists of 75 galaxy clusters that are covered by the EMU pilot survey in the redshift range of 0.03 < $z$ < 1.1. Ten clusters are cool-core clusters, while the rest are non-cool-core clusters. In 64 clusters we could identify a radio source that corresponds to the cluster BCG. The radio luminosities of the central radio galaxies at 944 MHz range between $\sim 10^{29}$ and $\sim 10^{33}$~erg s$^{-1}$ Hz$^{-1}$. The X-ray luminosities of the corresponding clusters range between $\sim 10^{43}$ and $\sim 10^{45}$ erg s$^{-1}$. 
      
      \item[] We compared the offset of the BCG from the cluster center to its concentration parameter. We found a link between the dynamical state of the cluster to its BCG offset. Larger offsets were found in more strongly disturbed systems.
      
      \item[] We find a statistically significant correlation between the radio and the X-ray luminosity, as in previous work \citep{Mittal_2009, Pasini_2020, Pasini_2021, Pasini_2022}. %This correlation can be interpreted as an
      %indication for the trend that most massive clusters host the most powerful radio-AGN.

      \item[] We investigated the correlation between the LLS of the radio source and its radio power, finding that larger radio galaxies tend to be more powerful. We do not find a correlation between the central density and the LLS, which suggests that the radio power is more important than the ambient density for the size of the radio galaxy.
      
      \item[] The 944 MHz luminosities were converted into 1.4 GHz kinetic luminosities using scaling relations. We found that in high-luminosity clusters with $L_{\mathrm{X}} > 10^{43}$ erg s$^{-1}$ , the kinetic luminosity of the radio jets is not longer correlated with the X-ray luminosity. This indicates that the variability in the AGN population is higher in high luminous clusters.
      
      \item[] We found an anticorrelation between the central cooling time $t_{\mathrm{cool}}$ and the radio luminosity $L_{\mathrm{R}}$ , indicating that more powerful AGN reside in clusters with short central cooling times.
      
      \item[] The density profiles of the individual clusters show that cool-core clusters tend to host powerful radio sources, in contrast to non-cool-core clusters, which host both high- and low-luminosity radio sources. 
      
      \item[] A mid-infrared color criterion using WISE colors was applied to our sample. We conclude that the color criterion is not applicable for our sample, which is due to the luminosity range of our sample, for which the WISE criterion is not applicable.
      
   \end{itemize}
   
The eRASS cluster catalog is a powerful tool that will prove useful for future studies. The combination with  radio surveys by the forthcoming generation of radio telescopes will vastly extend samples such as ours. 

\begin{acknowledgements}
The authors thank the anonymous referee for useful comments and suggestions.\\
The Australian SKA Pathfinder is part of the Australia Telescope National Facility which is managed by CSIRO. Operation of ASKAP is funded by the Australian Government with support from the National Collaborative Research Infrastructure Strategy. ASKAP uses the resources of the Pawsey Supercomputing center. Establishment of ASKAP, the Murchison Radio-astronomy Observatory and the Pawsey Supercomputing center are initiatives of the Australian
Government, with support from the Government of Western Australia and the Science and Industry Endowment Fund. We acknowledge the Wajarri Yamatji people as the traditional owners of the Observatory site. The Australia Telescope Compact Array (/ Parkes radio tele-
scope / Mopra radio telescope / Long Baseline Array) is part of the Australia Telescope National Facility which is funded by the Australian Government for operation as a National Facility managed by CSIRO. This paper includes archived data obtained through the Australia Telescope Online Archive (http://atoa.atnf.csiro.au). This work was supported by resources provided by the Pawsey Supercomputing center with funding from the Australian Government
and the Government of Western Australia. We acknowledge and thank the builders of ASKAPsoft.  \\
This work is based on data from eROSITA, the soft X-ray instrument aboard SRG, a joint Russian-German science mission supported by the Russian Space Agency (Roskosmos), in the interests of the Russian Academy of Sciences represented by its Space Research Institute (IKI), and the Deutsches Zentrum für Luft- und Raumfahrt (DLR). The SRG spacecraft was built by Lavochkin Association (NPOL) and its subcontractors, and is operated by NPOL with support from the Max Planck Institute for Extraterrestrial Physics (MPE). The development and construction of the eROSITA X-ray instrument was led by MPE, with contributions from the Dr. Karl Remeis Observatory Bamberg and ECAP (FAU Erlangen-Nuernberg), the University of Hamburg Observatory, the Leibniz Institute for Astrophysics Potsdam (AIP), and the Institute for Astronomy and Astrophysics of the University of Tübingen, with the support of DLR and the Max Planck Society. The Argelander Institute for Astronomy of the University of Bonn and the Ludwig Maximilians Universität Munich also participated in the science preparation for eROSITA. \\
This publication makes use of data products from the Wide-field Infrared Survey Explorer, which is a joint project of the University of California, Los Angeles, and the Jet Propulsion Laboratory/California Institute of Technology, funded by the National Aeronautics and Space Administration. \\
KB and MB are funded by the Deutsche Forschungsgemeinschaft (DFG, German Research Foundation) under Germany’s Excellence Strategy – EXC 2121 „Quantum Universe“ – 390833306.
\end{acknowledgements}

\bibliographystyle{aa}
\bibliography{Bib}

\clearpage
\onecolumn

\begin{appendix}
\section{eRASS:1 clusters in the EMU field}
We show a table of all eRASS:1 detected clusters in the EMU field and their corresponding X-ray characteristics. The radio values of the corresponding central radio source are also listed.
\begin{landscape}
\begin{longtable}{lrrrrrrrrrl}
\toprule
      Cluster ID &      RA &     DEC &     z & S$_{\mathrm{Radio}}$ (mJy) & LLS (kpc) & L$_{\mathrm{X}}$ ($10^{43}$ erg s$^{-1}$) & M (10$^{14}$M$_{\odot}$) & r$_{500}$ & $n_e$ (cm$^{-3}$) &                           BCG ID \\
\midrule
\endfirsthead
\toprule
      Cluster ID &      RA &     DEC &     z & S$_{\mathrm{Radio}}$ (mJy) & LLS (kpc) & L$_{\mathrm{X}}$ ($10^{43}$ erg s$^{-1}$) & M (10$^{14}$M$_{\odot}$) & r$_{500}$ & $n_e$ (cm$^{-3}$) &                           BCG ID \\
\midrule
\endhead
\midrule
\multicolumn{11}{r}{{Continued on next page}} \\
\midrule
\endfoot

\bottomrule
\caption{eRASS:1 clusters in the EMU field and their corresponding values. RA, DEC and z refer to the galaxy clusters, S$_{\mathrm{Radio}}$ is the radio flux of the corresponding central radio source in mJy, LLS is the largest linear size of the radio source in kpc, L$_{\mathrm{X}}$ is the X-ray luminosity of the cluster in $10^{43}$ erg s$^{-1}$, M is the cluster mass in 10$^{14}$M$_{\odot}$, r$_{500}$ is listed in kpc, $n_e$ is the central electron density of the cluster in cm$^{-3}$ and BCG ID is the identifier for the clusters brightest central galaxy.}
\endlastfoot
J201601.3-495436 & 304.006 & -49.910 & 0.271 &                     28.652 &       120 &                          15.562 &                    4.654 &      1075 &             0.067 &            J201601.62-495445.5 \\
J201847.9-524238 & 304.700 & -52.711 & 0.050 &                    290.843 &        50 &                           5.826 &                    2.581 &       953 &             0.361 &          2MASX J20184669-5241274 \\
J202154.3-525715 & 305.476 & -52.954 & 0.139 &                    142.582 &        88 &                           8.768 &                    3.295 &      1004 &             0.012 &              J202147.81-525705.2 \\
J202321.7-553524 & 305.840 & -55.590 & 0.231 &                     13.325 &       368 &                          38.238 &                    8.796 &      1349 &             0.005 & [GSB2009] J202320.83-553549.9  \\
J203328.8-593552 & 308.370 & -59.598 & 0.200 &                    353.543 &       126 &                          12.290 &                    4.099 &      1057 &             0.005 &              J203331.05-593549.4 \\
J202555.3-511709 & 306.481 & -51.286 & 0.229 &                     25.188 &       216 &                          28.201 &                    7.165 &      1260 &             0.005 &       2MASX J20255579-5116276  \\
J202726.9-522215 & 306.862 & -52.371 & 0.064 &                      5.976 &        22 &                           1.624 &                    1.066 &       707 &             0.013 &          6dFGS gJ202728.4-522210 \\
J203601.2-513931 & 309.005 & -51.659 & 0.272 &                    107.958 &       112 &                           7.305 &                    2.763 &       903 &             0.017 &        WISEA J203557.84-513909.5 \\
J203220.5-562738 & 308.086 & -56.461 & 0.284 &                      0.750 &        77 &                          39.745 &                    8.832 &      1324 &             0.005 &    [GSB2009] J203223.81-562759.4 \\
J203043.7-563749 & 307.682 & -56.630 & 0.394 &                     27.148 &       143 &                          24.071 &                    5.940 &      1113 &             0.002 &    [BRS2016] J203045.25-563755.8 \\
J204612.1-575544 & 311.551 & -57.929 & 0.225 &                    308.326 &       133 &                           5.985 &                    2.445 &       882 &             0.043 &        WISEA J204611.70-575550.0 \\
J203826.6-561509 & 309.611 & -56.253 & 0.368 &                     28.870 &       185 &                           3.910 &                    1.701 &       741 &             0.006 &              J203825.76-561522.3 \\
J205146.0-604621 & 312.942 & -60.773 & 0.337 &                      3.043 &        89 &                          25.940 &                    6.387 &      1165 &             0.037 &              J205145.73-604623.1 \\
J204408.5-603931 & 311.035 & -60.659 & 0.121 &                      2.018 &        42 &                           3.832 &                    1.882 &       838 &             0.003 &       2MASX J20441046-6039212  \\
J204008.3-503254 & 310.035 & -50.549 & 0.149 &                     54.060 &       121 &                           7.783 &                    3.020 &       972 &             0.074 &          2MASX J20401004-5032544 \\
J205555.5-545538 & 313.981 & -54.927 & 0.140 &                     21.865 &       187 &                          14.443 &                    4.660 &      1127 &             0.120 &          2MASX J20555594-5455493 \\
J205156.7-523751 & 312.987 & -52.631 & 0.045 &                      4.952 &        16 &                           1.838 &                    1.166 &       732 &             0.018 &                       ESO 187-26 \\
J204822.8-613113 & 312.095 & -61.520 & 0.108 &                      0.750 &       126 &                           7.862 &                    3.086 &       993 &             0.006 &          2MASX J20482154-6131025 \\
J205316.0-620912 & 313.317 & -62.154 & 0.395 &                      0.313 &       109 &                          13.545 &                    3.978 &       973 &             0.011 &              J205315.09-620906.2 \\
J210114.4-554134 & 315.310 & -55.693 & 0.260 &                      0.750 &        75 &                          19.559 &                    5.486 &      1140 &             0.002 &                      LEDA 406911 \\
J205943.1-501908 & 314.930 & -50.319 & 0.331 &                      0.763 &        84 &                           7.361 &                    2.659 &       872 &             0.002 &              J205941.29-501810.3 \\
J211652.8-593039 & 319.220 & -59.511 & 0.058 &                     60.742 &        32 &                           4.347 &                    2.089 &       886 &             0.018 &                           FRL 95 \\
J210604.1-584425 & 316.517 & -58.740 & 1.126 &                      8.465 &       382 &                         202.092 &                   17.422 &      1195 &             0.014 &     [FAB2011] 316.50647-58.73848 \\
J211250.8-531753 & 318.212 & -53.298 & 0.223 &                      8.623 &        65 &                           9.300 &                    3.328 &       978 &             0.062 &                      LEDA 431554 \\
J212023.5-542845 & 320.098 & -54.479 & 0.241 &                      5.831 &       149 &                           7.038 &                    2.720 &       909 &             0.016 &              J212025.34-542840.9 \\
J210732.3-552840 & 316.885 & -55.478 & 0.349 &                      2.027 &       128 &                          12.747 &                    3.898 &       984 &             0.016 &              J210732.69-552821.0 \\
J211144.8-533856 & 317.937 & -53.649 & 0.443 &                      3.612 &       113 &                          15.140 &                    4.190 &       972 &             0.011 &              J211144.61-533852.8 \\
J212251.3-582948 & 320.714 & -58.497 & 0.293 &                      0.750 &        82 &                          10.150 &                    3.425 &       962 &             0.015 &        WISEA J212249.41-582941.9 \\
J213151.8-500345 & 322.966 & -50.063 & 0.457 &                      0.750 &       209 &                          21.909 &                    5.376 &      1051 &             0.018 &              J213151.23-500344.2 \\
J213003.2-504832 & 322.514 & -50.809 & 0.076 &                     46.760 &        40 &                           6.662 &                    2.794 &       970 &             0.001 &          2MASX J21294244-5049260 \\
J212809.3-484330 & 322.039 & -48.725 & 0.321 &                      0.750 &       142 &                           5.134 &                    2.101 &       809 &             0.009 &            J212809.39-484346.8 \\
J213800.9-600758 & 324.504 & -60.133 & 0.319 &                      5.875 &        85 &                          54.931 &                   10.906 &      1402 &             0.003 &  [RBB2014] J213800.82-600753.8 \\
J212433.5-612500 & 321.140 & -61.417 & 0.436 &                     17.086 &       127 &                          34.158 &                    7.389 &      1177 &             0.006 &    [RBB2014] J212434.77-612444.4 \\
J213305.7-594531 & 323.274 & -59.759 & 0.505 &                      0.258 &       252 &                          22.120 &                    5.268 &      1024 &             0.007 &              J213305.59-594537.1 \\
J213221.5-585412 & 323.090 & -58.903 & 0.496 &                      2.215 &       129 &                          15.786 &                    4.192 &       952 &             0.022 &              J213221.89-585414.1 \\
J214528.9-513623 & 326.371 & -51.606 & 0.054 &                      0.750 &        18 &                           1.333 &                    0.937 &       679 &             0.318 &          2MASX J21452953-5136250 \\
J213505.3-625454 & 323.772 & -62.915 & 0.223 &                      4.264 &       103 &                           5.903 &                    2.482 &       887 &             0.005 &                      LEDA 338113 \\
J214359.4-563717 & 325.998 & -56.622 & 0.082 &                    168.404 &        48 &                          12.907 &                    4.377 &      1125 &             0.570 &                     MRC 2140-568 \\
J214622.3-571714 & 326.593 & -57.287 & 0.073 &                      4.689 &        41 &                           9.799 &                    3.631 &      1060 &             0.243 &                     FAIRALL 0116 \\
J214553.0-564448 & 326.471 & -56.747 & 0.481 &                      2.758 &       182 &                          51.416 &                    9.632 &      1264 &             0.007 &    [RBB2014] J214551.96-564453.5 \\
J213536.8-572622 & 323.903 & -57.440 & 0.427 &                      4.632 &       216 &                          28.547 &                    6.568 &      1136 &             0.008 &  [RBB2014] J213537.41-572630.7 \\
J214647.5-573648 & 326.698 & -57.614 & 0.611 &                      1.018 &       164 &                          33.250 &                    6.585 &      1058 &             0.035 &    [BRS2016] J214648.41-573653.7 \\
J214758.4-572019 & 326.993 & -57.339 & 0.142 &                      0.750 &       107 &                           5.916 &                    2.521 &       917 &             0.002 &                                0 \\
J215813.2-485044 & 329.555 & -48.846 & 0.499 &                      0.750 &       109 &                           7.569 &                    2.499 &       800 &             0.014 &              J215812.61-485031.7 \\
J215129.7-552019 & 327.874 & -55.339 & 0.041 &                   2623.450 &        80 &                           1.761 &                    1.150 &       730 &             0.074 &          2MASX J21512991-5520124 \\
J220153.8-595644 & 330.474 & -59.946 & 0.107 &                      2.129 &        48 &                          47.838 &                   10.776 &      1507 &             0.038 &                        ESO 146-5 \\
J220504.4-592716 & 331.268 & -59.455 & 0.350 &                      2.647 &       154 &                          22.796 &                    5.845 &      1126 &             0.010 &              J220500.42-592716.9 \\
J215444.0-593638 & 328.684 & -59.611 & 0.429 &                      2.036 &       147 &                          23.542 &                    5.736 &      1085 &             0.003 &              J215446.49-593634.0 \\
J214844.5-611650 & 327.185 & -61.281 & 0.572 &                      2.545 &       158 &                          37.403 &                    7.294 &      1111 &             0.003 &    [RBB2014] J214838.82-611555.9 \\
J215826.3-602359 & 329.610 & -60.400 & 0.085 &                     17.604 &        54 &                           5.648 &                    2.494 &       932 &             0.005 &          2MASX J21582577-6023286 \\
J220052.4-515502 & 330.219 & -51.917 & 0.218 &                      6.362 &       112 &                          10.970 &                    3.746 &      1019 &             0.011 &                      LEDA 451041 \\
J215625.0-513110 & 329.105 & -51.520 & 0.494 &                      1.940 &       116 &                          24.325 &                    5.657 &      1053 &             0.034 &              J215624.79-513119.4 \\
J220756.6-522333 & 331.986 & -52.393 & 0.111 &                      2.340 &        40 &                           3.082 &                    1.619 &       800 &             0.010 &              J220757.72-522320.2 \\
J215815.7-502328 & 329.566 & -50.391 & 0.488 &                      0.750 &       106 &                          25.476 &                    5.886 &      1069 &             0.010 &              J215817.77-502352.7 \\
J220604.7-494313 & 331.520 & -49.720 & 0.128 &                      5.430 &        62 &                           1.284 &                    0.883 &       650 &             0.018 &          2MASX J22060802-4943077 \\
J215919.0-521025 & 329.829 & -52.174 & 0.498 &                      3.824 &       151 &                          17.470 &                    4.498 &       974 &             0.005 &              J215919.21-520953.1 \\
J215515.7-514004 & 328.815 & -51.668 & 0.228 &                      0.750 &        65 &                           1.404 &                    0.903 &       632 &             0.659 &                2dFGRS TGS823Z290 \\
J215406.4-575136 & 328.527 & -57.860 & 0.075 &                      1.363 &        29 &                          22.077 &                    6.373 &      1278 &             0.005 &          2MASX J21540421-5752033 \\
J215918.0-564510 & 329.825 & -56.753 & 0.279 &                      0.750 &        76 &                          15.875 &                    4.684 &      1074 &             0.335 &        WISEA J215918.14-564459.1 \\
J221033.1-570945 & 332.638 & -57.163 & 0.300 &                      0.925 &        83 &                          11.120 &                    3.636 &       979 &             0.005 &                2dFGRS TGS812Z361 \\
J220952.8-553521 & 332.470 & -55.589 & 0.168 &                    135.047 &       137 &                           3.865 &                    1.858 &       821 &             0.008 &                      LEDA 407921 \\
J221117.3-483401 & 332.822 & -48.567 & 0.257 &                      1.026 &        96 &                          27.300 &                    6.899 &      1232 &             0.004 &                      LEDA 486024 \\
J220731.3-492529 & 331.881 & -49.425 & 0.254 &                     13.086 &       101 &                           5.273 &                    2.238 &       847 &             0.006 &              J220733.00-492444.0 \\
J220112.5-614737 & 330.302 & -61.794 & 0.238 &                     10.265 &        85 &                          13.025 &                    4.166 &      1048 &             0.126 &                      LEDA 352331 \\
J221631.8-532501 & 334.133 & -53.417 & 0.180 &                     83.864 &        80 &                           4.536 &                    2.064 &       847 &             0.008 &                      LEDA 430211 \\
J221454.0-532017 & 333.725 & -53.338 & 0.338 &                    144.263 &       210 &                           5.068 &                    2.072 &       800 &             0.007 &              J221454.87-532100.6 \\
J220920.7-514811 & 332.336 & -51.803 & 0.117 &                      0.791 &        40 &                          12.732 &                    4.321 &      1107 &             0.001 &                      LEDA 163750 \\
J222327.0-522740 & 335.863 & -52.461 & 0.273 &                     22.105 &       194 &                          12.071 &                    3.891 &      1012 &             0.008 &        WISEA J222321.82-522749.3 \\
J222000.0-522730 & 335.000 & -52.459 & 0.106 &                      0.439 &        32 &                           7.729 &                    3.055 &       990 &             0.000 &          2MASX J22200853-5227489 \\
J222120.9-501707 & 335.337 & -50.286 & 0.179 &                    600.143 &       129 &                           5.556 &                    2.376 &       888 &             0.010 &                    LEDA 468032 \\
J221504.0-520501 & 333.767 & -52.084 & 0.500 &                      2.117 &       148 &                          26.143 &                    5.939 &      1068 &             0.003 &                      LEDA 448827 \\
J222420.5-503901 & 336.085 & -50.650 & 0.335 &                      0.750 &       153 &                           8.557 &                    3.018 &       908 &             0.008 &              J222417.04-503859.3 \\
J221959.2-482901 & 334.997 & -48.484 & 0.871 &                      0.750 &       131 &                          81.107 &                   10.560 &      1116 &             0.008 &                                0 \\
J222251.8-483423 & 335.716 & -48.573 & 0.652 &                      0.476 &       119 &                          68.961 &                   11.090 &      1238 &             0.007 &    [BRS2016] J222250.73-483435.5 \\
J221959.1-581546 & 334.996 & -58.263 & 0.281 &                      6.828 &        90 &                           7.639 &                    2.826 &       907 &             0.002 &              J221958.01-581620.9 \\
\end{longtable}

\end{landscape}
\end{appendix}

\end{document}